\documentclass[english]{article}
\usepackage[T1]{fontenc}
\usepackage[latin9]{inputenc}
\usepackage{geometry}
\usepackage{float}
\usepackage{amsmath}
\usepackage{amssymb}
\usepackage{graphicx}
\usepackage{wasysym}

\makeatletter

\providecommand{\tabularnewline}{\\}
\floatstyle{ruled}
\newfloat{algorithm}{tbp}{loa}
\providecommand{\algorithmname}{Algorithm}
\floatname{algorithm}{\protect\algorithmname}

\newcommand{\lyxaddress}[1]{
	\par {\raggedright #1
	\vspace{1.4em}
	\noindent\par}
}

\@ifundefined{date}{}{\date{}}
\makeatother

\usepackage{babel}
\begin{document}

\title{A fully implicit, scalable, conservative nonlinear relativistic Fokker-Planck
0D-2P solver for runaway electrons }

\author{Don Daniel\thanks{corresponding author email: dond@lanl.gov}$\hspace{1mm}$,
William T. Taitano, Luis Chacón}
\maketitle

\lyxaddress{\begin{center}
Los Alamos National Laboratory, New Mexico, USA
\par\end{center}}
\begin{abstract}
Upon application of a sufficiently strong electric field, electrons
break away from thermal equilibrium and approach relativistic speeds.
These highly energetic \textquoteleft runaway\textquoteright{} electrons
(\ensuremath{\sim} MeV) play a significant role in tokamak disruption
physics, and therefore their accurate understanding is essential to
develop reliable mitigation strategies. For this purpose, we have
developed a fully implicit solver for the 0D-2P (i.e., including two
momenta coordinates) relativistic nonlinear Fokker-Planck equation
(rFP). As in earlier implicit rFP studies (NORSE, CQL3D), electron-ion
interactions are modeled using the Lorentz operator, and synchrotron
damping using the Abraham-Lorentz-Dirac reaction term. However, our
implementation improves on these earlier studies by 1) ensuring exact
conservation properties for electron collisions, 2) strictly preserving
positivity, and 3) being scalable algorithmically and in parallel.
Key to our proposed approach is an efficient multigrid preconditioner
for the linearized rFP equation, a multigrid elliptic solver for the
Braams-Karney potentials {[}Braams and Karney, \emph{Phys. Rev. Lett.}
\textbf{59}, 16 (1987){]}, and a novel adaptive technique to determine
the associated boundary values. We verify the accuracy and efficiency
of the proposed scheme with numerical results ranging from small electric-field
electrical conductivity measurements to the accurate reproduction
of runaway tail dynamics when strong electric fields are applied.
\end{abstract}

\section{Introduction}

Relativistic Coulomb collisions are modeled using an extended version
of the Landau-Fokker-Planck collision operator \cite{BB55}. Similarly
to its non-relativistic counterpart \cite{Rosenbluth}, the operator
assumes small-angle collisions, is well-posed, and features strict
conservation of total particle number, total momentum, and total energy
\cite{BK89}. However, its accurate numerical solution is difficult
because of its integro-differential formulation, which introduces
scalability and discretization challenges. In this study, we propose
a finite-difference-based conservative, parallel, fully implicit solver
for the 0D-2P relativistic Fokker-Planck (rFP) electron-electron collision
operator. This work builds and improves on earlier rFP algorithms
as implemented in the NORSE \cite{NORSE} and CQL3D \cite{Harvey2000,CQL3D}
codes.

The solver proposed in this study is primarily designed to simulate
runaway electrons produced by a large loop voltage. In tokamaks, the
loop voltage induced during disruptions can produce a large amount
of runaway electrons, which may severely damage plasma facing materials
\cite{Woods,CH75}. The generated runaway current is also affected
by secondary mechanisms such as energy transfers from the primary
runaway electron current to the thermal electrons through knock-on
(large-angle) collisions. Understanding these nonlinear mechanisms
may be essential to develop either avoidance or mitigation strategies
for runaway electrons in tokamaks. 

A solver designed to capture runaway-electron dynamics benefits from
certain features. For example, capturing small-amplitude tails necessitates
strict positivity preservation. Runaway-electron generation time may
be large: a sizable runaway tail length may take hundreds of electron-electron
thermal collision-time scales to develop. Therefore, an implicit solver
that can step over stiff thermal collision-time scales is desirable.
The ability to use large time steps also demands that the scheme be
asymptotic preserving, which in turn requires enforcing strict conservation
properties \cite{WT2015}. It is also essential that the solver be
optimal and scales with the number of mesh points, as resolving small-scale
features may require fine grids while fitting long tails may require
large domains. The relativistic Fokker-Planck operator can be expressed
either in integral form \cite{BB55} or in differential form \cite{BK87}.
We employ the differential form, in which the collisional coefficients
are expressed in terms of relativistic potentials. This form is more
conducive to an optimal $\mathcal{{O}}(N)$ solver, where $N$ is
the number of grid points, as the integral form produces an $\mathcal{{O}}(N^{2})$
scaling when symmetry is preserved in the integral operators (which
is required to achieve strict conservation properties \cite{hirvijoki2019conservative,shiroto2019structure}).
To obtain a nearly optimal scaling ($\mathcal{{O}}(N^{\alpha}\log N)$,
with $\alpha\gtrsim1$) with the differential formulation, we propose
a multigrid-preconditioned GMRES \cite{saad1986gmres} solver for
the potentials along with an adaptive treatment for evaluating potential
far-field boundary conditions {[}which greatly decreases their computational
complexity from $\mathcal{O}(N^{3/2})$ to $\mathcal{O}(N^{1.1})${]}.
We note that it is in principle possible to improve the scaling resulting
from the integral form of the collision operator by the use of optimal
integral methods such as fast multipole methods \cite{fmm}. However,
such methods can break the numerical symmetry of the integrals, resulting
in the loss of strict conservation properties. In this regard, fast
integral methods would not improve on the proposed optimal algorithm
for the differential formulation, and would require similar strategies
to recover the strict conservation properties of the collision operator.

With regard to time-stepping, we propose a conservative, fully implicit
nonlinear scheme, which as a result is asymptotic preserving (i.e.,
it captures the Maxwell-Jüttner distribution as a steady-state solution
to our system). Earlier algorithmic approaches proposed for this system
are either linearly implicit (e.g., NORSE \cite{NORSE}), or lack
strict conservation properties (e.g., CQL3D \cite{Harvey2000,CQL3D}).
The implicit solver proposed in this study satisfies discrete conservation
properties, is preconditioned for optimal algorithmic performance,
and is scalable in parallel. As a result, our algorithm scales as
$\mathcal{{O}}(N^{1.1}\log N).$ Our conservation and preconditioning
strategies follow closely those proposed in Ref. \cite{WT2015}.

The rest of the paper is organized as follows. In $\S$\ref{sec:Formulation},
we discuss the full relativistic electron-electron operator, the Lorentz
operator for electron-ion interactions, and the Abraham-Lorentz-Dirac
reaction term for modeling losses due to synchrotron damping. Then,
in $\S$\ref{sec:Algorithm-1}, we discuss the algorithmic aspects
with regard to the discrete conservation strategy, positivity preservation,
and our near-optimal strategy for determination of the potentials.
In $\S$\ref{sec:Nonlinear-Solver-Strategy-1}, we briefly describe
our fully implicit nonlinear solver using an Anderson Acceleration
scheme. In $\S$\ref{sec:Numerical-results}, we discuss the numerical
results that demonstrate the correctness of our implementation, and
finally in $\S$\ref{sec:Conclusion-and-Future} we list the conclusions
and scope for future work. 

\section{Formulation\label{sec:Formulation}}

We model a homogeneous quasi-neutral plasma. We evolve the electron
species with the relativistic Fokker-Planck equation for the electron
distribution function, $f_{e}$, in the presence of background species
$\beta$, 

\begin{equation}
\partial_{t}f_{e}+\partial_{\vec{{p}}}\cdot\left[({\vec{{E}}}+\vec{{F}}_{S})f_{e}\right]=\sum_{\beta=i,e}C(f_{\beta},f_{e}),\label{eq:RFP-1}
\end{equation}
where $t$ is time normalized with the relativistic electron collision
time, 
\[
\tau_{ee}^{relativistic}=\frac{{4\pi\epsilon_{0}^{2}m_{e}^{2}c^{3}}}{q_{e}^{4}n_{e}\ln\Lambda_{ee}},
\]
$\vec{{p}}$ is the momentum vector normalized with $m_{e}c$, $m_{e}$
is the electron mass, $c$ is the speed of light, $q_{e}$ is the
electron charge, $\vec{{E}}$ is the electric field normalized with
the critical value for runaway electron generation \cite{CH75}, $E_{c}=n_{e}q_{e}^{3}\ln\Lambda_{ee}/4\pi\epsilon_{0}^{2}m_{e}c^{2}$,
$n_{e}$ is the electron number density, $\epsilon_{0}$ is the electrical
permittivity, $\ln\Lambda_{ee}$ is the Coulomb logarithm, $\vec{{F}}_{S}$
refers to the electron friction coefficients associated with synchrotron
radiation damping effects (defined in detail later), and $C$ is the
collision operator given by, 

\begin{equation}
C(f_{\beta},f_{e})=\partial_{\vec{{p}}}\cdot\left[\overline{{\overline{{D}}}}_{\beta}\cdot\partial_{\vec{{p}}}f_{e}-\frac{{m_{e}}}{m_{\beta}}\vec{{F_{\beta}}}f_{e}\right],\label{eq:collision-1-1}
\end{equation}
where $\overline{{\overline{{D}}}}_{\beta}$ represents the collisional
diffusion tensor coefficients and $\vec{{F}}_{\beta}$ represents
the collisional friction vector coefficients (computed based on the
appropriate background species $f_{\beta}$). Though Eq. (\ref{eq:RFP-1})
in principle may be used for multiple species, here we only consider
the evolution of electrons interacting with themselves, ions and external
electric fields. 

\begin{figure}
$\qquad$\includegraphics[viewport=0bp 0bp 800bp 218bp,scale=0.45]{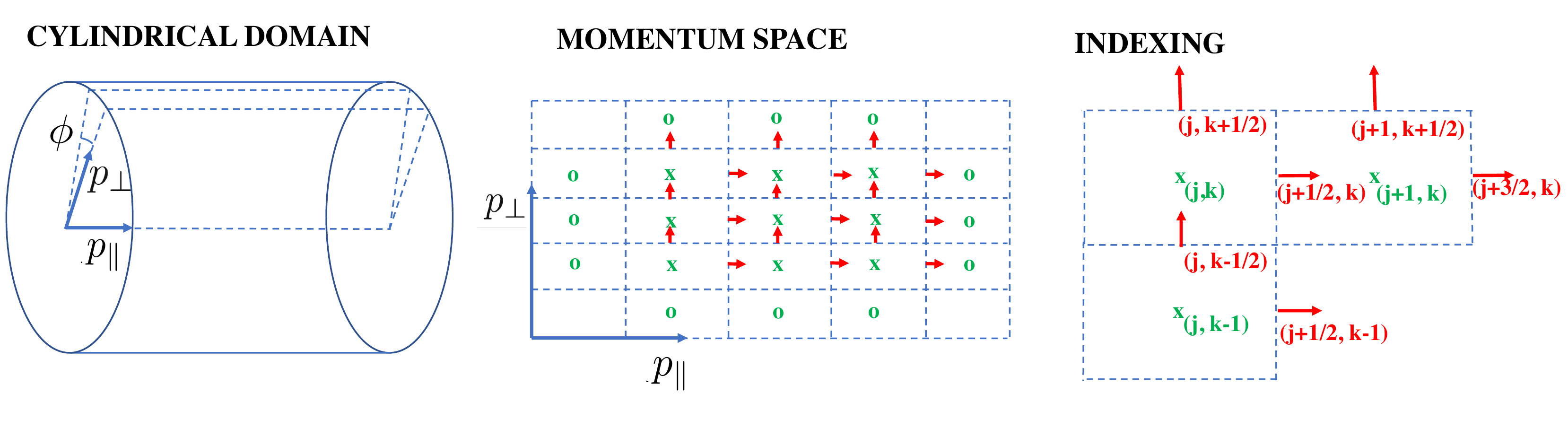}

\caption{We consider a cylindrical geometry representation $(p_{\parallel,}p_{\perp})$
with azimuthal symmetry (left). Following a finite volume formulation,
we define the distribution function on cell centers (crosses) and
fluxes on edges (arrows). The ghost cells (circles) are exterior to
domain boundaries. A typical stencil is shown on the right. The discrete
volume for cell $(j,k)$ is computed as $\Delta V_{j,k}=2\pi p_{\perp,k}\Delta p_{\parallel,j}\Delta p_{\perp,k},$
where $\Delta p_{\parallel,j}$ and $\Delta p_{\perp,k}$ are the
discrete momentum space cell sizes in the parallel and perpendicular
directions.\label{fig:domain} }
\end{figure}

The distribution function is described in a two-dimensional cylindrical
domain $(p_{\parallel},p_{\perp}$), with the subscripts $\parallel$
and $\perp$ referring to directions parallel and perpendicular to
the magnetic field, respectively (see Fig. \ref{fig:domain}). The
azimuthal direction is ignored because the distribution is axisymmetric.
The electron-electron interactions are described using the full form
of the collision operator, while the electron-ion interaction is modeled
with the Lorentz operator (which assumes the ions to be cold and infinitely
massive, $m_{i}>>\gamma m_{e}$ with $\gamma=\sqrt{{1+p^{2}}}$ the
Lorentz factor). 

\subsection{Electron-electron collisions}

The collisional coefficients, $\overline{{\overline{{D}}}}_{\beta}$
and $\vec{{F}}_{\beta}$, for electrons in Eq. (\ref{eq:collision-1-1})
are expressed in terms of the Braams-Karney potentials \cite{BK87}.
These potentials are obtained by inverting a set of elliptic equations.
In this study, the elliptic solves are performed optimally with parallel
multigrid-preconditioned GMRES techniques, with a scaling of $\mathcal{{O}}(N\log N)$.
The collisional coefficients are given by \cite{BK87}:

\begin{eqnarray}
\overline{{\overline{{D}}}}_{e} & = & -\frac{{4\pi}}{n_{\beta}}\gamma^{-1}[\overline{{\overline{{L}}}}+\overline{{\overline{{P}}}}]h_{1}+4\gamma^{-1}[\overline{{\overline{{L}}}}-\overline{{\overline{{P}}}}]h_{2},\label{eq:D-1}\\
\vec{{F}_{e}} & = & -\frac{{4\pi}}{n_{\beta}}\gamma^{-1}\vec{{K}}(g_{0}-2g_{1}),\label{eq:A-1}
\end{eqnarray}
where the operators $\overline{{\overline{{L}}}}$, $\vec{{K}}$,
and $\overline{{\overline{{P}}}}$ are defined as:
\begin{eqnarray*}
\overline{{\overline{{L}}}}\psi & = & \overline{{\overline{{P}}}}\cdot\frac{{\partial^{2}\psi}}{\partial{\vec{{p}}}\partial{\vec{{p}}}}\cdot\overline{{\overline{{P}}}}-\overline{{\overline{{P}}}}\left(\vec{{p}}\cdot\frac{{\partial{\psi}}}{\partial\vec{{p}}}\right),\\
\vec{{K}}\psi & = & \overline{{\overline{{P}}}}\cdot\frac{{\partial{\psi}}}{\partial{\vec{{p}}}},\\
\overline{{\overline{{P}}}} & = & \overline{{\overline{{I}}}}+\vec{{p}}\vec{{p}}.
\end{eqnarray*}
To obtain the transport coefficients, we first compute the $h$ potentials
by solving the partial differential equations, 

\begin{eqnarray}
[L+1]h_{0} & = & f_{e},\nonumber \\{}
[L-3]h_{1} & = & h_{0},\label{eq:hpotentials-1}\\{}
[L-3]h_{2} & = & h_{1},\nonumber 
\end{eqnarray}
and then the $g$ potentials by solving: 

\begin{eqnarray}
Lg_{0} & = & f_{e},\nonumber \\
Lg_{1} & = & g_{0}.\label{eq:gpotentials-1}
\end{eqnarray}
Here, the operator $L$ is defined as:
\begin{equation}
L\,\psi=\overline{{\overline{{P}}}}:\frac{{\partial^{2}\psi}}{\partial{\vec{{p}}}\partial{\vec{{p}}}}+3\vec{{p}}\cdot\frac{{\partial{\psi}}}{\partial\vec{{p}}}.\label{eq:potentialoperator}
\end{equation}
To solve these linear potential equations, we require far-field boundary
conditions. They are determined from the Green's function solution
of the elliptic equations, Eqs. (\ref{eq:hpotentials-1},\ref{eq:gpotentials-1})
\cite{BB55}:

\begin{eqnarray}
h_{0} & = & -\frac{{1}}{4\pi}\int(r^{2}-1)^{-1/2}\frac{{f_{\beta}(\vec{{p'}})}}{\gamma'}\mathrm{{d}^{3}\vec{{p}'},}\nonumber \\
h_{1} & = & -\frac{{1}}{8\pi}\int\sqrt{{r^{2}-1}}\frac{{f_{\beta}(\vec{{p'}})}}{\gamma'}\mathrm{{d}^{3}\vec{{p}'},}\nonumber \\
h_{2} & = & -\frac{{1}}{32\pi}\int(r\cosh^{-1}{r}-\sqrt{{r^{2}-1}})\frac{{f_{\beta}(\vec{{p'}})}}{\gamma'}\mathrm{{d}^{3}\vec{{p}'},}\label{eq:potentialGreen-1}\\
g_{0} & = & -\frac{{1}}{4\pi}\int r(r^{2}-1)^{-1/2}\frac{{f_{\beta}(\vec{{p'}})}}{\gamma'}\mathrm{{d}^{3}\vec{{p}'},}\nonumber \\
g_{1} & = & -\frac{{1}}{8\pi}\int\cosh^{-1}{r}\frac{{f_{\beta}(\vec{{p'}})}}{\gamma'}\mathrm{{d}^{3}\vec{{p}'},}\nonumber 
\end{eqnarray}
where $r=\gamma\gamma'-\vec{{p}}\cdot\vec{{p'}}$. Note that the integral
kernels of $h_{0}$ and $g_{0}$ are singular when $r\rightarrow1\,(\vec{{p}}\rightarrow\vec{{p'}}),$
which require a specialized numerical treatment in terms of elliptic
integrals for accuracy and efficiency (see $\S$\ref{subsec:collisionalpotentials}
and App. \ref{app:potential_elliptic_kernel}). Also, we have devised
an efficient adaptive algorithm to fill ghost cells that prevents
these boundary integrals from leading to an $\mathcal{{O}}(N^{3/2})$
scaling of the computational complexity (see also $\S$\ref{subsec:collisionalpotentials}).

\subsection{Modeling external effects\label{subsec:Modeling-external-effects}}

We consider several external effects, including an imposed electric
field, $\vec{{E}}=(E_{\parallel},0)$, ions, and synchrotron radiation,
$\vec{{F}}_{S}$. 

Electron-ion scattering is modeled with the Lorentz or pitch-angle
scattering operator \cite{BK89,decker2016numerical,NORSE}, which
assumes ions are cold and infinitely massive. The operator causes
scattering of the electrons in the pitch angle $(\arccos[p_{\parallel}/p])$
and, in this simplified form, it preserves kinetic energy. It has
finite diffusion coefficients and zero friction coefficients, given
by:

\begin{equation}
D_{i,\parallel\parallel}=\frac{{Z_{\mathrm{{eff}}}}}{2v}\frac{{p_{\perp}^{2}}}{p^{2}},\qquad D_{i,\parallel\perp}=D_{i,\perp\parallel}=-\frac{{Z_{\mathrm{{eff}}}}}{2v}\frac{{p_{\perp}p_{\parallel}}}{p^{2}},\qquad D_{i,\perp\perp}=\frac{{Z_{\mathrm{{eff}}}}}{2v}\frac{{p_{\parallel}^{2}}}{p^{2}},\qquad\vec{{F}}_{i}=\vec{{0}},\label{eq:electron-ion-scattering}
\end{equation}
where $p^{2}=p_{\perp}^{2}+p_{\parallel}^{2}$, $v$ is the velocity
magnitude (normalized with $c$), and $Z_{\mathrm{{eff}}}=\sum n_{i}Z_{i}^{2}/\sum n_{i}Z_{i}$
is the effective ion-charge state ($n_{i}$ and $Z_{i}$ refers to
ion densities and charges). For a quasi-neutral plasma, $\sum n_{i}Z_{i}=n_{e}.$
Note that the electron-ion collision operator becomes singular at
the origin $v\rightarrow0$. We mollify this singularity by reformulating
the singular part, as:

\[
\frac{{1}}{v}\approx\frac{{1}}{\sqrt{{v^{2}+v_{cut}^{2}}}},
\]
where $v_{cut}=p_{cut}/\sqrt{{1+p_{cut}^{2}}}$ is the velocity cut-off,
with $p_{cut}=2\Delta p$. Note this approximation of the singular
term in the cylindrical space introduces a finite but small amount
of heating as $p\rightarrow0$. 

Finally, we consider synchrotron radiation, which results in loss
of momentum for the electrons. We model this with the Abraham-Lorentz-Dirac
reaction term \cite{decker2016numerical,Guo17}. The reaction term
has finite friction coefficients, given by: 

\begin{equation}
F_{S,\perp}=-S\frac{{p_{\perp}}}{\gamma}(1+p_{\perp}^{2}),\qquad F_{S,\parallel}=-S\frac{{p_{\parallel}}}{\gamma}p_{\perp}^{2},\label{eq:synchrotron_damping}
\end{equation}
where $S=\tau_{ee}^{relativistic}/\tau_{r}$ relates the time scale
of the synchrotron-radiation damping, $\tau_{r}$, to that of electron-electron
relativistic collisions, $\tau_{ee}^{relativistic}$.

\section{Algorithm\label{sec:Algorithm-1}}

\subsection{General discretization strategy}

We employ a conservative finite-difference scheme. The distribution
is evaluated at cell centers, while the friction and diffusion fluxes
are evaluated at cell faces. Recall the electron-electron collision
operator is the divergence of a collisional flux,

\begin{equation}
C(f_{e},f_{e})=\partial_{\vec{{p}}}\cdot(\overline{{\overline{D}}}_{e}\nabla_{p}f_{e}-\vec{{F_{e}}}f_{e})\approx\delta_{\vec{{p}}}\cdot(\vec{{R}}_{D}-\vec{R}_{F})=\delta_{\vec{{p}}}\cdot\vec{{R}},\label{eq:FPflux}
\end{equation}
where $\vec{{R}}_{D}$ and $\vec{{R}}_{F}$ are the diffusion and
friction fluxes, respectively, and $\delta_{\vec{{p}}}\cdot$ denotes
the discrete form of the divergence operator, which in cylindrical-momentum
space is written as:

\begin{equation}
\left(\delta_{\vec{{p}}}\cdot\vec{{R}}\right)_{j,k}=\left(\frac{{R_{\parallel,j+1/2,k}-R_{\parallel,j-1/2,k}}}{\Delta p_{\parallel,j}}+\frac{{p_{\perp,k+1/2}R_{\perp,j,k+1/2}-p_{\perp,k-1/2}R_{\perp,j,k-1/2}}}{p_{\perp,k}\Delta p_{\perp,k}}\right).\label{eq:fluxformulation}
\end{equation}
Fluxes at cell faces are given by:

\[
R_{D,\parallel,j+\frac{{1}}{2},k}=\left(D_{\parallel\parallel}\partial_{p_{\parallel}}f_{e}+D_{\parallel\perp}\partial_{p_{\perp}}f_{e}\right)_{j+\frac{1}{2},k},\quad R_{D,\perp,j,k+\frac{{1}}{2}}=\left(D_{\perp\parallel}\partial_{p_{\parallel}}f_{e}+D_{\perp\perp}\partial_{p_{\perp}}f_{e}\right)_{j,k+\frac{{1}}{2}},
\]

\[
R_{F,\parallel,j+\frac{{1}}{2},k}=F_{\parallel,j+\frac{1}{2},k}f_{e,j+\frac{1}{2},k},\quad R_{F,\perp,j,k+\frac{{1}}{2}}=F_{\perp,j+\frac{1}{2},k}f_{e,j,k+\frac{{1}}{2}}.
\]

The potential operator $L$ (Eq. \ref{eq:potentialoperator}) is discretized
using central differences (see App. A1 for details). The potentials
are evaluated at cell centers and their boundary conditions are specified
at ghost cells. For the potential Eqs. (\ref{eq:hpotentials-1})-(\ref{eq:gpotentials-1}),
we apply far-field Dirichlet boundary conditions using Eqs. (\ref{eq:potentialGreen-1})
as discussed in $\S$\ref{subsec:collisionalpotentials}. The collisional
coefficients, $\overline{{\overline{{D}}}}_{\beta}$ and $\vec{{F}}_{\beta}$,
are evaluated at cell centers using the computed potentials (see App.
A2 for discretization details). The collisional coefficients at the
ghost cells are evaluated by linearly extrapolating the values from
adjacent cell-centered values. Note that the ghost cells also store
distribution and coefficient data for cross-processor communication
using an MPI framework. 

The coefficients for external effects (such as scattering due to ion
interactions, Eq. (\ref{eq:electron-ion-scattering}), synchrotron
damping effects, Eq. (\ref{eq:synchrotron_damping}), and electric
field acceleration terms) are evaluated at cell-centers. Where needed,
values at cell faces are found by averaging two adjacent cell-centered
values within the computational domain. As is typical in kinetic simulations,
the outer domain boundaries are selected such that the distribution
is sufficiently small there.

\subsection{Discrete conservation strategy for the e-e collision operator}

\label{subsec:Discrete-conservation-strategy}

The relativistic electron-electron collision operator conserves the
total particle number, momentum, $\vec{{p}}=\gamma\vec{{v}}$, and
energy $E=\gamma$, as the moments of the collision operator satisfy:

\begin{eqnarray}
\langle1,C(f_{e},f_{e})\rangle_{p} & = & 0,\label{eq:massconservation}\\
\langle p_{\parallel},C(f_{e},f_{e})\rangle_{p} & = & 0,\label{eq:pconservation}\\
\langle\gamma,C(f_{e},f_{e})\rangle_{p} & = & 0,\label{eq:Econservation}
\end{eqnarray}
where $\langle a,b\rangle_{p}=\int_{p}ab2\pi p_{\perp}\mathrm{{d}}p_{\parallel}\mathrm{{d}}p_{\perp}$.
Discretely, these inner products may be approximated by a mid-point
quadrature rule as: 

\[
\langle A,B\rangle_{p}^{D}\approx2\pi\sum_{j=1}^{N_{\parallel}}\sum_{k=1}^{N_{\perp}}A_{j,k}B_{j,k}p_{\perp,k}\Delta p_{\parallel,j}\Delta p_{\perp,k},
\]
where the superscript $D$ refers to the discrete representation of
the summation operator, and $\Delta p_{\parallel,j}$ and $\Delta p_{\perp,k}$
are the width and height of a rectangular cell located at $(j,k).$

In general, the relationships shown in Eqs. (\ref{eq:massconservation}-\ref{eq:Econservation})
will not be satisfied due to numerical errors. Discrete particle number
conservation (Eq. \ref{eq:massconservation}) is trivially satisfied
by setting the normal component of diffusion and friction fluxes to
zero at the boundary. However, enforcing Eqs. (\ref{eq:pconservation},\ref{eq:Econservation})
is more challenging. A recent study \cite{WT2015} enforced these
conservation properties discretely by redistributing the numerical
errors via discrete nonlinear constraints. We employ a similar methodology
here. Firstly, we multiply the diffusion flux by a factor

\[
\eta=1+\eta_{0}+\eta_{1}(p_{\parallel}-\bar{{p}}_{\parallel}),
\]
where the magnitudes of $\eta_{0}$ and $\eta_{1}$ are expected to
be of the order of truncation error, and $\bar{{p}}_{\parallel}=\langle f_{e},p_{\parallel}\rangle_{p}/\langle1,f_{e}\rangle_{p}$
is the mean momentum. Thus, the discrete collision operator is of
the form,

\[
C^{D}(f_{e},f_{e})=\delta_{\vec{{p}}}\cdot(\vec{\eta{R_{D}}}-\vec{R}_{F}),
\]
Integrating over the cylindrical-momentum domain, 

\begin{eqnarray}
\langle p_{\parallel},C^{D}(f_{e},f_{e})\rangle_{p}^{D} & = & 0,\nonumber \\
\langle\gamma,C^{D}(f_{e},f_{e})\rangle_{p}^{D} & = & 0,\label{eq:discretesum}
\end{eqnarray}
we obtain a system of two equations,

\begin{equation}
\left[\begin{array}{cc}
\langle\gamma\delta_{\vec{{p}}}\cdot\vec{{R}}\rangle_{p}^{D} & \langle\gamma(p_{\parallel}-\bar{{p}}_{\parallel})\delta_{\vec{{p}}}\cdot\vec{{R}}_{D}\rangle_{p}^{D}\\
\langle p_{\parallel}\delta_{\vec{{p}}}\cdot\vec{{R}}_{D}\rangle_{p}^{D} & \langle p_{\parallel}(p_{\parallel}-\bar{{p}}_{\parallel})\delta_{\vec{{p}}}\cdot\vec{{R}}_{D}\rangle_{p}^{D}
\end{array}\right]\left[\begin{array}{c}
\eta_{0}\\
\eta_{1}
\end{array}\right]=\left[\begin{array}{c}
\langle\gamma\delta_{\vec{{p}}}\cdot(\vec{{R}}_{F}-\vec{{R}}_{D})\rangle_{p}^{D}\\
\langle p_{\parallel}\delta_{\vec{{p}}}\cdot(\vec{{R}}_{F}-\vec{{R}}_{D})\rangle_{p}^{D}
\end{array}\right],\label{eq:discreteconseq}
\end{equation}
for unknowns $[\eta_{0},\:\eta_{1}]$, which can be inverted straightforwardly.
This strategy conserves momentum and energy at the discrete level
for electron-electron collisions. Note that, because we assume the
ions to be cold and infinitely massive, there are no conservation
properties associated with electron-ion collisions (i.e. discrete
representation may result in a net energy and momentum loss).

\subsection{Time stepping strategy\label{subsec:Time-stepping-strategy}}

A huge separation in time scales exists in runaway-electron dynamics.
Long time-scales are of the order of the relativistic collision times,
$\mathcal{{O}}(\tau_{ee}^{relativistic})$. For typical bulk temperatures
$\Theta=T/m_{e}c^{2}\sim10^{-4}$, this implies a time-scale separation
of six orders of magnitude between thermal and relativistic time scales
(as $\tau_{ee}^{thermal}=\Theta^{3/2}\tau_{ee}^{relativistic}$).
Stepping over fast time scales demands a fully implicit temporal scheme
with strict conservation and positivity preservation properties. We
describe our approach next.

The discrete system of equations representing the effects of electron-electron
collisional interactions $C$ and external effects, $\mathcal{{E}}$,
on electron evolution can be written as:

\begin{equation}
\delta_{t}f_{e}^{n}=C^{D}(f_{e}^{n},f_{e}^{n})+\underbrace{{\tilde{{C}}(f_{i,}^{n}f_{e}^{n})-\delta_{\vec{{p}}}.\left[(\vec{{F}}_{S}+\vec{{E}})f_{e}^{n}\right]}}_{\mathcal{{E}}(f_{e}^{n})},\label{eq:timeadvancement}
\end{equation}
where the superscript $D$ represents the appropriate discrete form
defined in $\S$\ref{subsec:Discrete-conservation-strategy}, and
$\tilde{{C}}$ represents the Lorentz operator (see Eq. \ref{eq:electron-ion-scattering}).
For a general implicit backward time discretization scheme at time
step $n$, we have,

\[
\delta_{t}f_{e}^{n}=\sum_{i=0,1,2...}\frac{{b_{i}f_{e}^{n-i}}}{\Delta t},
\]
where constants, $b_{i}$, satisfy $\sum_{i}b_{i}=0.$ We use both
first-order (Euler, BDF1) and second-order (BDF2) schemes for time
advancing. For BDF1, $b_{0}=-1$ and $b_{1}=1$ and for BDF2 with
constant time steps, $b_{o}=3/2,$ $b_{1}=-2$, $b_{2}=1/2$. The
coefficients can be generalized for non-uniform time steps. 

Multiplying Eq. (\ref{eq:timeadvancement}) with $\vec{{c}}=(1,p_{\parallel},\gamma$),
and averaging over the momentum space, we obtain: 

\[
\sum_{i=0,1,2...}\frac{{b_{i}\langle\vec{{c}},f_{e}^{n-i}\rangle_{p}^{D}}}{\Delta t}=\langle\vec{{c}},C^{D}(f_{e}^{n},f_{e}^{n})\rangle_{p}^{D}+\langle\vec{{c}},\mathcal{{E}}(f_{e}^{n})\rangle_{p}^{D}.
\]
Because of the discrete conservation properties of the electron-electron
collision operator, the first term in the right hand side vanishes.
Therefore, any change in the total momentum or energy of electrons
can only be due to external effects such as ion-electron collisions,
synchrotron radiation, and electric field acceleration. 

\subsection{Positivity-preserving strategy}

Positivity-preserving schemes are essential to capture small-amplitude
runaway tails. Our strategy is to leverage the structure of the differential
operators (advection-diffusion), and use existing positivity-preserving
discretizations for these terms.

For all advective terms in the relativistic kinetic equation, we use
the positivity-preserving SMART flux limiter \cite{SMART} to construct
the associated fluxes. For the diagonal components of the tensor diffusion
term, $\overline{{\overline{{D}}}}\cdot\nabla_{p}f_{e}\,|_{\parallel\parallel}$
and $\overline{{\overline{{D}}}}\cdot\nabla_{p}f_{e}\,|_{\perp\perp}$,
we employ a standard second-order discretization:

\[
(\overline{{\overline{{D}}}}\cdot\nabla_{p}f_{e}\,)_{\parallel\parallel,j+\frac{{1}}{2},k}=\left(D_{\parallel\parallel}\partial_{p_{\parallel}}f_{e}\right)_{j+\frac{{1}}{2},k}\approx\frac{{D_{\parallel\parallel,j+1,k}+D_{\parallel\parallel,j,k}}}{2}\frac{{f_{e,j+1,k}-f_{e,j,k}}}{\Delta p_{\parallel,j+\frac{{1}}{2}}},
\]

\[
(\overline{{\overline{{D}}}}\cdot\nabla_{p}f_{e}\,)_{\perp\perp,j,k+\frac{{1}}{2}}=\left(D_{\perp\perp}\partial_{p_{\perp}}f_{e}\right)_{j,k+\frac{{1}}{2}}\approx\frac{{D_{\perp\perp,j,k+1}+D_{\perp\perp,j,k}}}{2}\frac{{f_{e,j,k+1}-f_{e,j,k}}}{\Delta p_{\perp,k+\frac{{1}}{2}}},
\]
which is numerically well-posed (does not feature a null space and
features a maximum principle). However, unless care is taken, the
off-diagonal diffusion tensor terms do not feature a discrete maximum
principle, resulting in loss of boundedness. To address this issue,
we reformulate the off-diagonal components as effective friction forces
as proposed in Ref. \cite{DOV}:

\begin{eqnarray}
(\overline{{\overline{{D}}}}\cdot\nabla_{p}f_{e}\,)_{\parallel\perp} & = & D_{\parallel\perp}\partial_{p_{\perp}}f_{e}=f_{e}\underbrace{D_{\parallel\perp}\partial_{p_{\perp}}\ln f_{e}}_{F_{\parallel}^{\mathrm{{eff}}}}=f_{e}F_{\parallel}^{\mathrm{{eff}}}=R_{F,\parallel}^{\mathrm{{eff}}},\nonumber \\
(\overline{{\overline{{D}}}}\cdot\nabla_{p}f_{e}\,)_{\perp\parallel} & =f_{e} & \underbrace{D_{\perp\parallel}\partial_{p\parallel}\ln f_{e}}_{F_{\perp}^{\mathrm{{eff}}}}=f_{e}F_{\perp}^{\mathrm{{eff}}}=R_{F,\perp}^{\mathrm{{eff}}}.\label{eq:Feff}
\end{eqnarray}
Once formulated as advective terms, we use flux-limiting advective
schemes (similar to the collisional friction terms) to calculate the
effective flux. Discretization details can be found in App. A3. 

\subsection{Strategy for evaluating boundary conditions of collision potentials\label{subsec:collisionalpotentials} }

The boundary conditions for the relativistic potential equations for
$h$ and $g$ are found using the integral formulations in Eq. (\ref{eq:potentialGreen-1}).
For the $h_{1},h_{2},g_{1}$ relativistic potentials, we use a trapezoidal-rule
numerical integration with 24 discrete points in the $\phi$ angle.
However, the kernels in $g_{0},h_{0}$ become singular in the limit
of $p'\rightarrow p\implies r\rightarrow1,$ complicating a direct
numerical integration. These complexities can be eliminated by reformulating
these integrals in terms of complete elliptic integrals. We begin
by noting that, because the distribution is axisymmetric, the 3D momentum-space
integration of the Green's function can be rewritten as a 2D momentum
space integration over the PDF and a 1D azimuthal angle integration
as:

\begin{eqnarray*}
h_{\beta,0} & = & -\frac{{1}}{4\pi}\int\frac{{f_{\beta}(p'_{\parallel},p'_{\perp})}}{\gamma'}p_{\perp}dp'_{\parallel}dp'_{\perp}\underbrace{{\int_{\phi}\frac{{1}}{(r^{2}-1)^{1/2}}d\phi}}_{I}=-\frac{{1}}{4\pi}\int\frac{{f_{\beta}(p'_{\parallel},p'_{\perp})}}{\gamma'}I(p_{\parallel},p{}_{\perp},p'_{\parallel},p'_{\perp})p_{\perp}dp'_{\parallel}dp'_{\perp},\\
g_{\beta,0} & = & -\frac{{1}}{4\pi}\int\frac{{f_{\beta}(p'_{\parallel},p'_{\perp})}}{\gamma'}p_{\perp}dp'_{\parallel}dp'_{\perp}\underbrace{{\int_{\phi}\frac{{r}}{(r^{2}-1)^{1/2}}d\phi}}_{H}=-\frac{{1}}{4\pi}\int\frac{{f_{\beta}(p'_{\parallel},p'_{\perp})}}{\gamma'}H(p_{\parallel},p{}_{\perp},p'_{\parallel},p'_{\perp})p_{\perp}dp'_{\parallel}dp'_{\perp},
\end{eqnarray*}
The segregated integrals $I$ and $H$ are then written in terms of
complete integrals of the first and third kind (see App. \ref{app:potential_elliptic_kernel}).

\begin{figure}
$\hspace{5cm}$\includegraphics[scale=0.6]{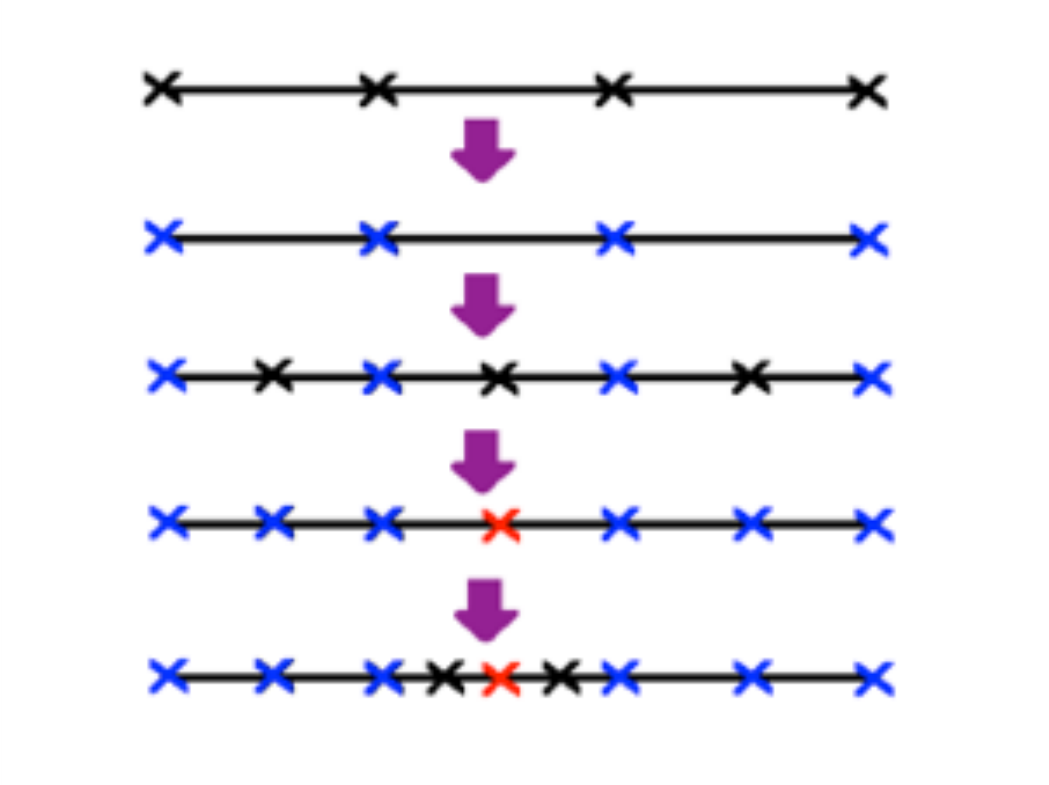}

\caption{Illustration of adaptive spline technique.\label{fig:Visualizing-adaptive-spline}}
\end{figure}

\begin{algorithm}
\begin{enumerate}
\item Initialize a set of knots. 
\item Evaluate potential integrals and create spline (cubic or higher-order). 
\item Bisect original knots to create new knots.
\item Evaluate potential integrals at each knot and check error using Eq.
(\ref{eq:boundaryerror}) : $\lvert{\phi_{I}-\phi_{S}}\rvert.$
\item Where error is small, stop local bisection. Where error is large,
go to step 3. 
\end{enumerate}
\caption{Adaptive spline based potential boundary treatment.}
\end{algorithm}
However, even after these reductions, evaluating potentials at all
ghost points in the boundary remains expensive. There are approximately
$\mathcal{{O}}(N^{1/2})$ ghost-cell boundary points, each point requiring
$\mathcal{{O}}(N)$ integrals when using Eqs. (\ref{eq:potentialGreen-1}).
This makes the potential boundary evaluations scale poorly with the
number of mesh points, $N$ {[}i.e., $\mathcal{{O}}(N^{3/2})${]}.
To ameliorate the scaling for the boundary-condition treatment, we
adaptively select a small number of boundary points for the potential
evaluations to match a given accuracy, with the remaining ghost points
found by interpolation using a high-order spline. The adaptive algorithm
to find the minimum number of spline knots needed for a given tolerance
is outlined in Algorithm 1, and illustrated in Fig. \ref{fig:Visualizing-adaptive-spline}.
We begin with a set of uniformly distributed ghost points at the boundary,
for example, four points (black crosses in the first row), where we
evaluate the values of the potential integral. We fit a cubic (or
higher order) spline through these values (blue crosses in the second
row). New knots are then created by bisection (black crosses in the
third row), where integrals are again evaluated. The absolute error
is then computed as the difference between the value given by the
spline interpolation, $\phi_{S}$, and the actual value of the potential
integral at the targeted points $\phi_{I}$:

\begin{equation}
a_{b}=\lvert{\phi_{I}-\phi_{S}}\rvert.\label{eq:boundaryerror}
\end{equation}
Intervals delimited by the set of knots that do not satisfy the prescribed
tolerance (e.g., red knot in the fourth row) are bisected further.
This process is continued until a spline fit of the desired accuracy
is obtained. 

\begin{figure}
\includegraphics[scale=0.61]{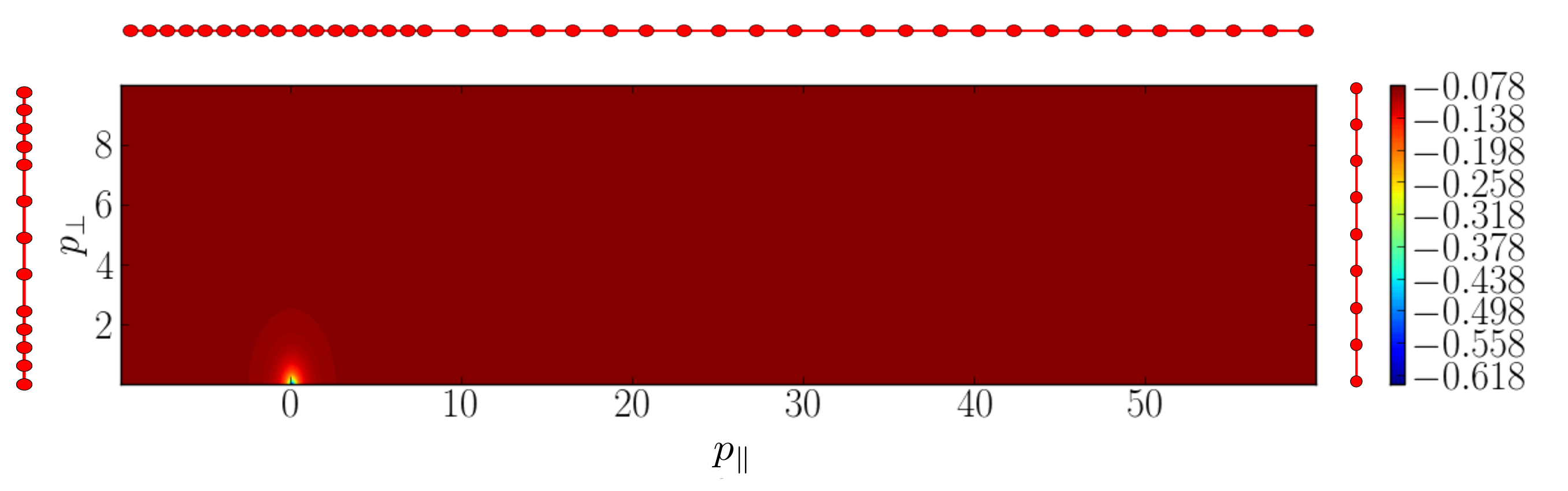}

\caption{Adaptive spline knots (fourth-order spline) in a uniform computational
domain with $N_{\parallel}=2048$ and $N_{\perp}=512.$ The figure
illustrates the $g_{0}$ potential for a Maxwell-Jüttner distribution
of $\Theta=10^{-2}$. The red dots represent the location of the spline
knots, comprising a total of 41 knots at the top boundary, and 13
and 9 knots at the left and right boundaries, respectively. \label{fig:Adaptive-spline-knots}}
\end{figure}
To ensure the spline error is commensurate with other sources of error
in the algorithm, in practice the absolute tolerance criterion is
chosen to be a function of the momentum mesh spacing as: 

\begin{equation}
a_{b}\sim0.05\Delta p_{\perp}\Delta p_{\parallel}.\label{eq:adaptive-spline-tol}
\end{equation}

Figure \ref{fig:Adaptive-spline-knots} illustrates the adaptive knots
generated with the adaptive spline algorithm for the $g_{0}$ potential
for a Maxwell-Jüttner distribution of $\Theta=10^{-2}$ in a mesh
of $N_{\parallel}=2048$ and $N_{\perp}=512.$ The red dots along
the left, top, and right boundaries point to the location of the spline
knots generated using Algorithm 1. The contour of $g_{0}$ is also
illustrated to demonstrate the variation of $g_{0}$ at points close
and far away from the distribution. The algorithm generates more spline
knots where the function varies significantly. At the far right boundary,
the points are few and equally spaced, as the function variation is
small. A clear benefit of the adaptive spline approach can be seen
at the top boundary, where it is determined that only 41 functional
evaluations are needed for an accurate estimate of the potential along
the entire boundary, which spans a total of $2048$ mesh points. 

Assuming equi-spaced knots, we can get an estimate of how the number
of splining knots, $N_{p}$, scales with the total degrees of freedom,
$N$, by comparing the spline error with the tolerance in Eq. \ref{eq:adaptive-spline-tol}:

\begin{equation}
\frac{{1}}{N}\sim\left(\frac{{1}}{N_{p}}\right)^{n_{s}+1}\implies N_{p}\propto N^{1/(n_{s}+1)}.\label{eq:knot-scaling}
\end{equation}
Here, $n_{s}$ is the order of the spine. For instance, for a fourth-order
spline, this result predicts $N_{p}\sim N^{0.2}$. However, we expect
this estimate to be very conservative, and it does not take into the
account the adaptive distribution of the knots. Numerical experiments
in Sec. $\S$\ref{subsec:Algorithmic-and-parallel} show that $N_{p}\sim\mathcal{O}(N^{0.13})$
for a fourth-order spline. 

\section{Nonlinear solver \label{sec:Nonlinear-Solver-Strategy-1}}

The spatial and temporal discretization techniques prescribed in $\S$\ref{sec:Algorithm-1}
lead to a coupled nonlinear system of equations, which requires an
iterative nonlinear solver for the distribution function. We use an
Anderson Acceleration scheme \cite{Anderson} to converge iteratively
the system, which we briefly summarize next. 

Given a fixed point map based Picard iteration,

\[
f^{k+1}=G(f^{k}),
\]
where the superscript $k$ denotes the iteration step, Anderson Acceleration
scheme \cite{walker2011anderson} accelerates the convergence of the
Picard iteration by using the history of past nonlinear solutions
via: 

\begin{equation}
f^{k+1}=\sum_{i=0}^{m_{k}}\alpha_{i}^{k}\underbrace{G(f^{k-m_{k}+i})}_{f^{k-m_{k}+i+1}},\label{eq:anderson}
\end{equation}
where in this study $m_{k}=\min(5,k).$ The coefficients $\alpha_{i}^{k}$
are determined via an optimization procedure that minimizes, 

\[
\left\Vert \sum_{i=0}^{m_{k}}\alpha_{i}^{k}\left(G(f^{k-m_{k}+i})-f^{k-m_{k}+i}\right)\right\Vert ,
\]
 subject to $\sum_{i=0}^{m_{k}}\alpha_{i}^{k}=1$.

To enable preconditioning of the Anderson iteration, our fixed map
is based on a quasi-Newton iteration, where:

\begin{equation}
f^{k+1}=G(f^{k})=f^{k}+\delta f^{k}=f^{k}-\left(P^{k}\right)^{-1}\mathcal{{R}}^{k},\label{eq:fixedpointmap}
\end{equation}
with $P^{k}$ the preconditioner, $\delta f^{k}$ the nonlinear increment,
and $\mathcal{{R}}^{k}$ the nonlinear residual. Given an electron
distribution, $f_{e}$, the residual for the nonlinear system is evaluated
as outlined in Algorithm 2. Note that if $P$ is the Jacobian, i.e.
$P^{k}=(\partial\mathcal{{R}}/\partial f_{e})^{k}$, then Eq. (\ref{eq:fixedpointmap})
becomes a Newton iteration. 

The residual contribution from electron-electron collisions requires
the solutions of five potentials, which require inversions of the
linear equations in Eqs. (\ref{eq:hpotentials-1},\ref{eq:gpotentials-1}).
These are inverted for each nonlinear iteration at flux-assembly time
along with the computation of conservation constraints $\eta_{0}$
and $\eta_{1}$, see Eq. \ref{eq:discreteconseq}. The nonlinear elimination
of the residuals associated with the potentials and conservation constraints
follows from previous studies \cite{Chacon2000,WT2015}, and enables
a conservative, optimal $\mathcal{{O}}(N\log N)$ solver when the
Poisson operators are inverted optimally and scalably. Here, the linear
potential equations are solved using a multigrid-preconditioned GMRES
\cite{GMRES} solver. The multigrid preconditioner features 1 V cycle
with 4 passes of damped Jacobi (damping factor of 0.7), along with
agglomeration for restriction and a second-order prolongation. At
the beginning of the solve, the five potentials are solved using a
tighter relative tolerance criteria of $10^{-8}$ and then followed
by a looser relative-tolerance criteria of $10^{-5}-10^{-7}$ during
each nonlinear solve, depending on the problem. 

The preconditioner in Eq. (\ref{eq:fixedpointmap}) is obtained by
Picard linearization of the potentials and subsequent discretization
of the full system,

\[
P^{k}\delta f=\delta_{t}\delta f-C(f_{e}^{k-1},\delta f)-\mathcal{{E}}(\delta f),
\]
where $\mathcal{{E}}$ is a linear operator representing the net external
effects on electrons, see Eq. (\ref{eq:timeadvancement}). The transport
coefficients in the electron-electron collision operator, $C$, are
Picard-linearized and computed at the previous nonlinear iteration,
$k$. All advective terms in the preconditioner are discretized using
a linear upwinding scheme. During each nonlinear step $k$, the linear
system $P^{k}\delta f^{k}=-\mathcal{{R}}^{k}$ is solved with one
multigrid V-cycle and 3 passes of damped Jacobi (with damping constant
0.7). We use agglomeration for restriction and second-order prolongation. 

The nonlinear iteration ends when the desired relative nonlinear residual
convergence ratio $r_{NL}$ is reached, 

\[
r_{NL}=\frac{\|{\mathcal{{R}}}^{k}\|}{\|{\mathcal{{R}}}^{k=0}\|}.
\]
Cases with large disparities in signal amplitudes, for example a Maxwellian
thermal bulk along with runaway tail, may require a tighter convergence
ratio, $r_{NL}=10^{-7}$, to capture accurately the small-amplitude
tail. In contrast, a single deforming electron thermal bulk can use
a significantly looser nonlinear convergence, $r_{NL}=10^{-4}$, for
accurate results.

\begin{algorithm}
\begin{enumerate}
\item Compute $\delta_{t}f_{e}$ and boundary conditions for potentials.
\item Invert potential equations for $h_{0},h_{1},h_{2},g_{0},g_{1}$ using
Eqs. (\ref{eq:hpotentials-1}-\ref{eq:gpotentials-1}) and evaluate
collisional coefficients.
\item Compute collision operator, $C(f_{e},f_{e})$, and enforce conservation
symmetries.
\item Compute external physics: electron-ion scattering operator, $\tilde{{C}}(f_{i,}f_{e})$,
synchrotron damping radiation and parallel electric field acceleration$,\,\delta_{\vec{{p}}}\cdot\left[(\vec{{F}}_{S}+\vec{{E}})f_{e}\right].$
\item Assemble nonlinear residual:
\[
\mathcal{{R}}(f_{e})=\delta_{t}f_{e}-C(f_{e},f_{e})-\tilde{{C}}(f_{i,}f_{e})+\delta_{\vec{{p}}}\cdot\left[(\vec{{F}}_{S}+\vec{{E}})f_{e}\right].
\]
\end{enumerate}
\caption{Evaluating nonlinear residual, $\mathcal{{R}}$ .}
\end{algorithm}

\section{Results\label{sec:Numerical-results}}

We begin this section with some verification studies, and finish it
with scalability and accuracy studies to assess the performance of
the algorithm.

\subsection{Verification}

We first verify conservation properties of the equilibrium Maxwell-Jüttner
distributions either at rest or moving with a mean momentum, $p_{b}.$
Note that all computations are performed in the stationary reference
frame. Then, we verify conservation properties during collisional
relaxation, and also benchmark the calculation of electrical conductivity
under the action of both weak and strong electric fields with previous
studies \cite{BK89,Weng,NORSE}. Finally, we verify our algorithm
with recent calculations of runaway dynamics in the $t\rightarrow\infty$
limit \cite{CH75,decker2016numerical,Guo17}. Verification results
were obtained using the second-order BDF2 time-stepping scheme except
when explicitly stated (see $\S$\ref{subsec:Time-stepping-strategy}
for details on the time-stepping scheme).

\subsubsection{Preservation of stationary and boosted Maxwell-Jüttner distributions. }

\begin{figure}
$(a)\hspace{8cm}(b)$

\includegraphics[scale=0.6]{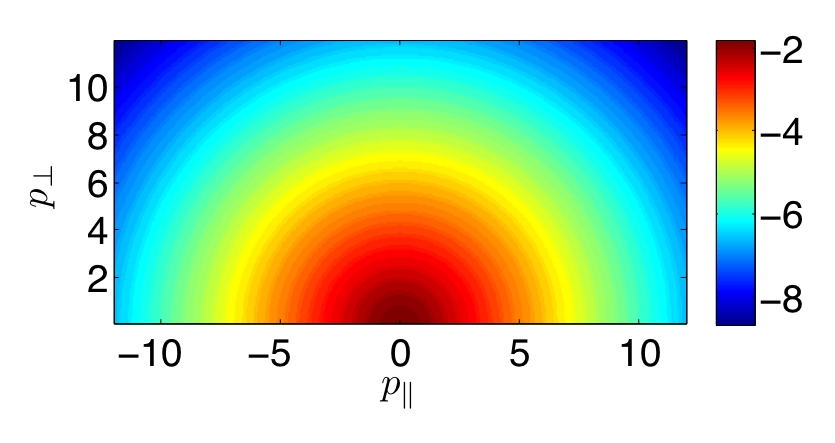}$\hspace{1cm}$\includegraphics[scale=0.6]{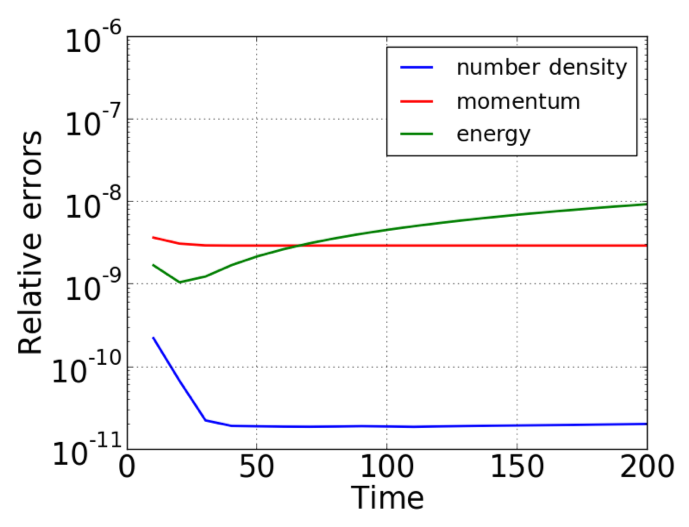}\caption{Preservation of a stationary Maxwell-Jüttner distribution, $f_{e}^{MJ}$,
for $\Theta=1$, $n_{e}=1,N_{\parallel}=256,$ and $N_{\perp}=128$,
see Eq. (\ref{eq:SMJ}). $(a)$ Log contour of electron distribution
$f_{e}$. The distribution remains unchanged as a function of time
(not shown). $(b)$ Time evolution of relative errors (Eq. \ref{eq:relerr})
of number density (blue), relativistic momentum (red) and relativistic
energy (green). Note time is normalized with $\tau_{ee}^{relativistic}$,
$\tau_{ee}^{relativistic}=\tau_{ee}^{thermal}$ as $\Theta=1$. }
\label{fig_ee_conservation}
\end{figure}

The computational domain is uniform with $N_{\parallel}=256$ and
$N_{\perp}=128.$ The nonlinear residual, $r_{NL}$, is converged
to a relative tolerance of $10^{-4}$ unless otherwise specified.
The discrete conservation properties are satisfied to nonlinear tolerance
and are independent of the time step used. The electron number density
is normalized, $n_{e}=1$. The domain is chosen such that the distribution
function is sufficiently small at boundaries. The entire domain is
shown in the figures illustrating the distribution function. Fig.
\ref{fig_ee_conservation}$(a)$ illustrates a static Maxwell-Jüttner
(MJ):

\begin{equation}
f_{e}^{MJ}=\frac{{n_{e}}}{4\pi\Theta K_{2}(1/\Theta)}\exp{\left[-\frac{{\gamma(p)}}{\Theta}\right]},\label{eq:SMJ}
\end{equation}
in log scale with normalized temperature $\Theta=T/m_{e}c^{2}=1$.
In Eq. \ref{eq:SMJ}, $K_{2}$ is the modified Bessel function of
the second kind. We have confirmed that the distribution retains its
initial shape for the whole simulation. To illustrate this, Fig. \ref{fig_ee_conservation}$(b)$
demonstrates the evolution of the relative errors in the number density,
momentum and energy for 200 $\tau_{ee}$. The relative errors are
measured as, 

\begin{equation}
\mathrm{{relative\,error}}=\frac{\lvert g(t)-g(0)\rvert}{g(0)}.\label{eq:relerr}
\end{equation}
where $g$ is either the number density, momentum or energy. The figure
shows that the relative errors in number density are one part in $10^{11}$,
while errors in relativistic momentum and energy remain smaller than
one part in $10^{8}$. 

\begin{figure}
$(a)\hspace{9cm}(b)$

\includegraphics[scale=0.6]{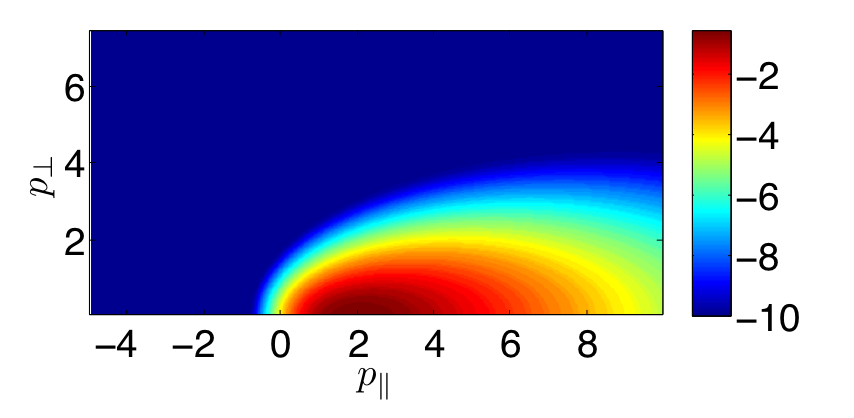}$\hspace{1cm}$\includegraphics[scale=0.6]{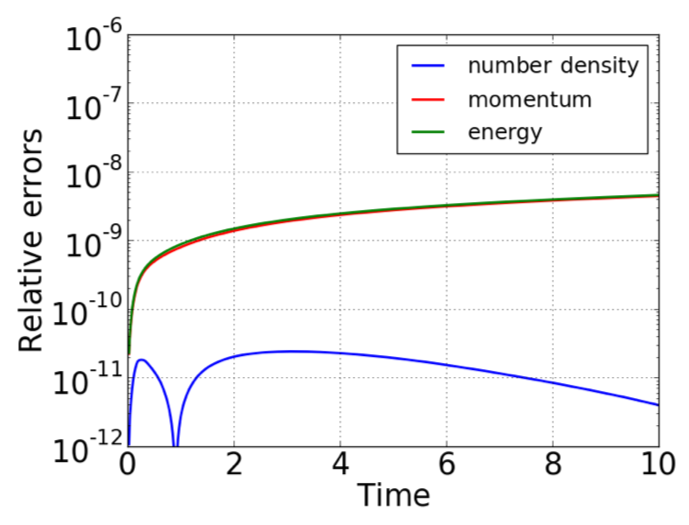}

\caption{Preservation of a boosted Maxwell-Jüttner distribution, $f_{e}^{BMJ}$,
for $n_{e}=1,\Theta_{b}=0.15,$ $p_{b}=2,N_{\parallel}=256,$ and
$N_{\perp}=128$, see Eq. \ref{eq:feBMJ}. $(a)$ Log contour of electron
distribution, $f_{e}$. $(b)$ Time evolution of relative errors in
number density, momentum, and energy. \label{fig_ee_boosted}}
\end{figure}

For a boosted (translated) MJ, the equilibrium distribution appears
deformed and is given by: 

\begin{equation}
f_{e}^{BMJ}=\frac{{n_{e}}}{4\pi\Theta_{b}\gamma_{b}K_{2}(1/\Theta_{b})}\exp{\left[-\frac{{\gamma_{b}\gamma-p_{b}p_{\parallel}}}{\Theta_{b}}\right]},\label{eq:feBMJ}
\end{equation}
where the subscript $b$ denotes the values in the boosted frame.
Fig. \ref{fig_ee_boosted} illustrates a boosted MJ equilibrium distribution
with $\Theta_{b}=0.15$, in a frame boosted by $p_{b}=2$ and with
$\gamma_{b}=\sqrt{{1+p_{b}^{2}}}.$ The relative errors are shown
for $10\tau_{ee}^{relativistic}\sim100\tau_{ee}^{thermal}$, demonstrating
identical behavior as in the stationary MJ case. 

\subsubsection{Conservation properties during collisional relaxation dynamics}

To explore collisional relaxation dynamics, we consider two cases,
one which features an initial configuration of two boosted MJ distributions,
and the other which features a randomized initial distribution. Simulations
have been run till the distributions have relaxed to a single Maxwell-Jüttner.

\begin{figure}
$(a)\hspace{7cm}(b)$

\includegraphics[scale=0.35]{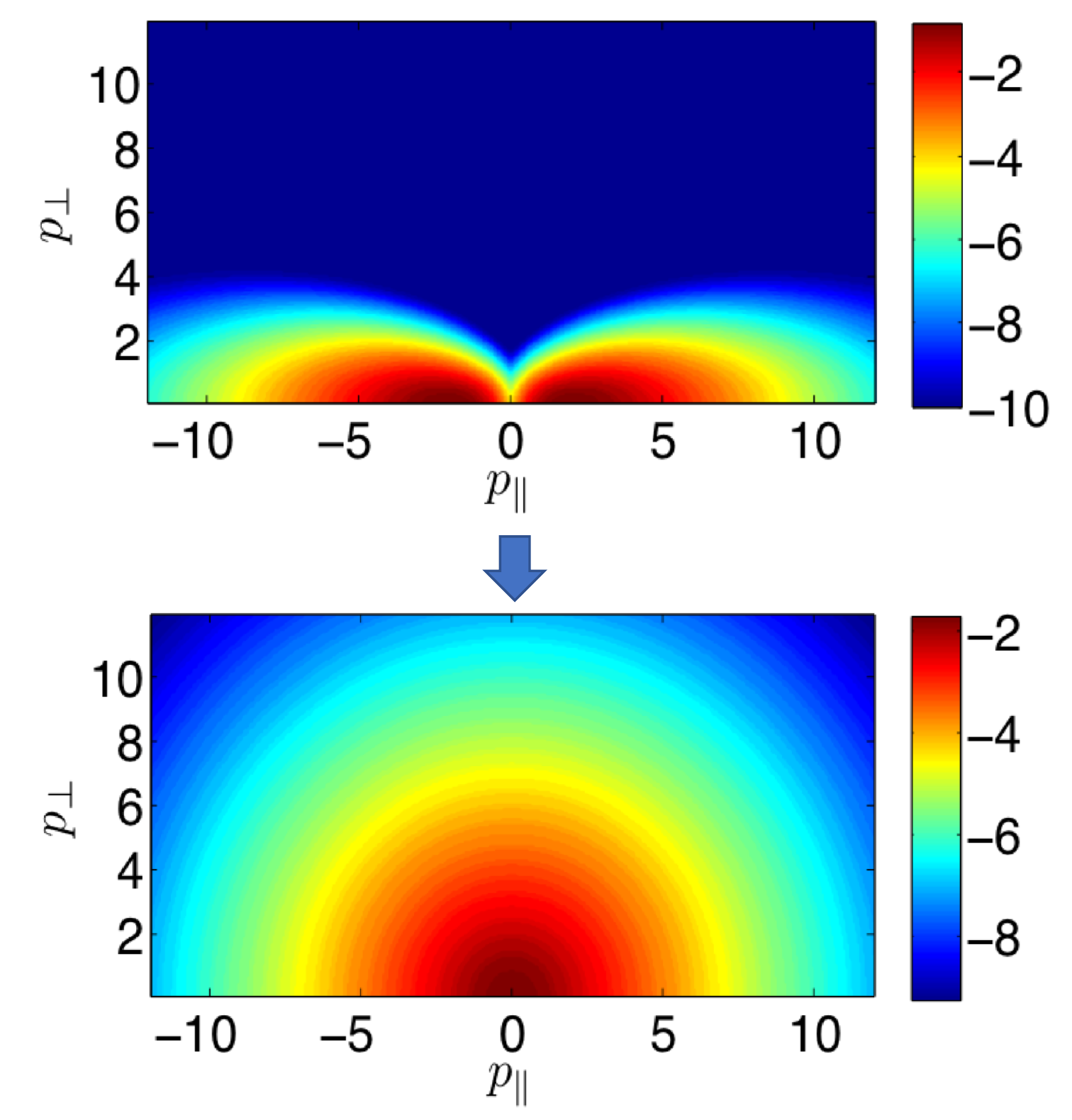}$\hspace{2cm}$\includegraphics[scale=0.6]{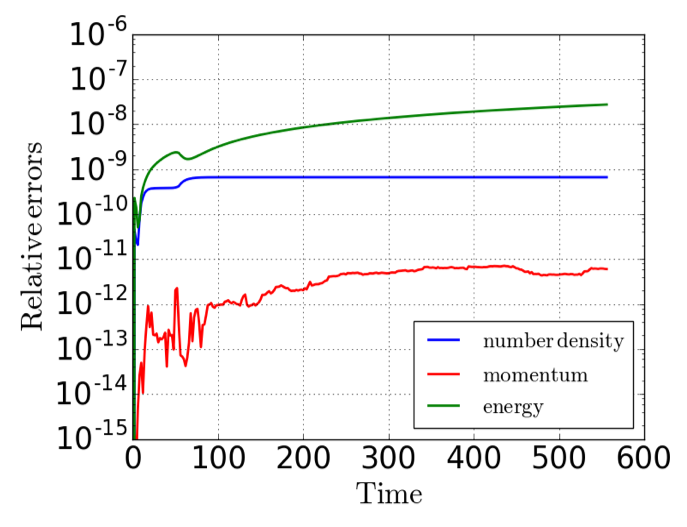}

\caption{Collisional relaxation of two boosted Maxwell-Jüttner distributions,
$f_{e}^{BMJ}$, with $n_{e}=1$, $p_{b}=-2,\,2$, $\Theta_{b}=0.15,N_{\parallel}=256,$
and $N_{\perp}=128$, see Eq. \ref{eq:feBMJ}. $(a)$ Evolution of
electron distribution contours from two distinct distributions at
initial time (top) to a single Maxwell-Jüttner at final time (bottom)
$(b)$ Time evolution of relative errors measured during the collisional
relaxation process. \label{fig_ee_ci_2MJ}}
\end{figure}

Figure \ref{fig_ee_ci_2MJ}$(a)$ illustrates the collisional relaxation
of two MJ distributions boosted by 2 units in opposite directions.
Fig. \ref{fig_ee_ci_2MJ}$(b)$ depicts the relative errors in number
density, momentum, and energy. After an initial transient stage $t\in(0,300)$,
the relative errors in momentum and number density remain small and
bounded in time. 

\begin{figure}
$(a)\hspace{5cm}\qquad(b)\hspace{5cm}(c)$

\includegraphics[width=6cm]{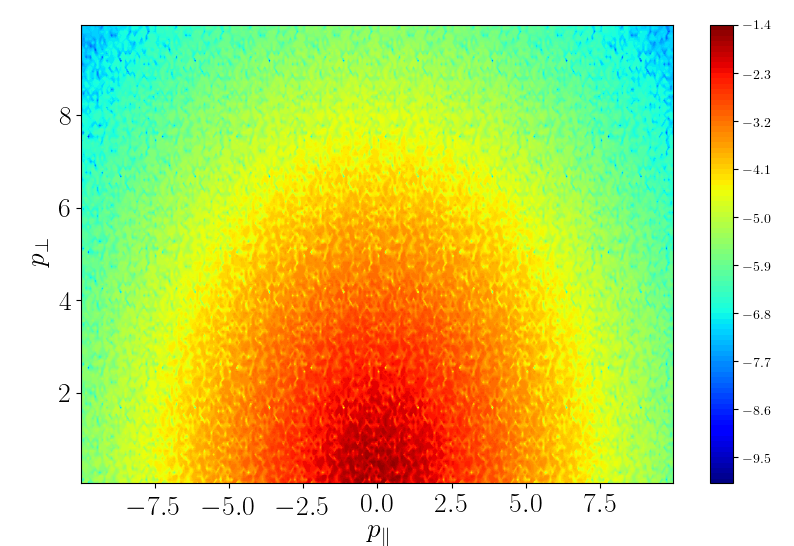}\includegraphics[scale=0.33]{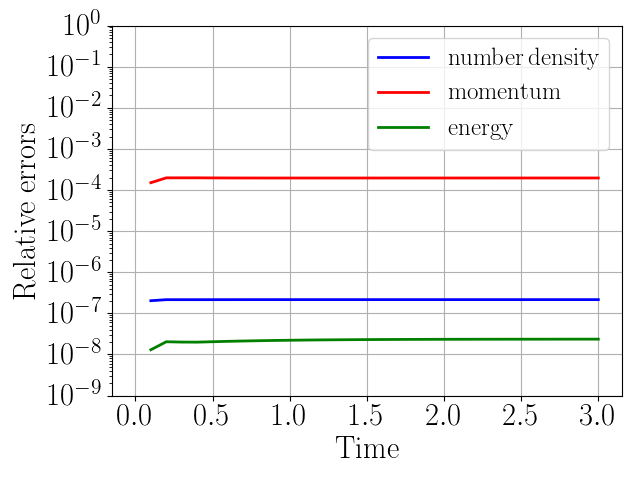}\includegraphics[scale=0.33]{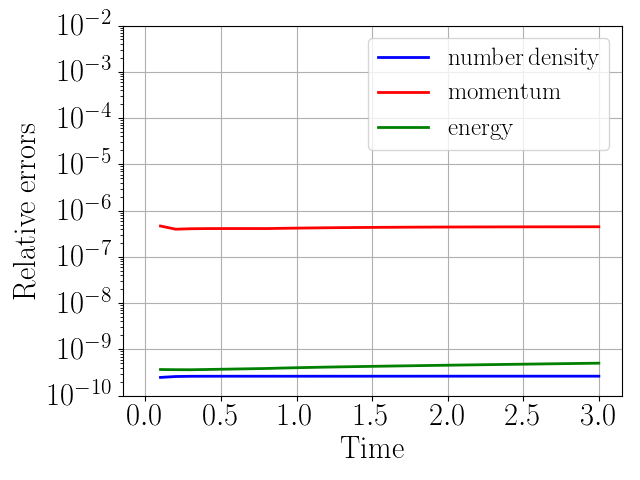}

\caption{Thermal relaxation of a random perturbed Maxwell-Jüttner distribution,
$f_{e}^{rand}$, for $N_{\parallel}=256$ and $N_{\perp}=128$, see
Eq. (\ref{eq:ferandom}). $(a)$ Initial random electron distribution,
see Eq. (\ref{eq:ferandom}). $(b)-(c)$ Evolution of relative errors
in discrete conservation properties for a nonlinear relative tolerance
of $10^{-4}$ in $(b)$ and $10^{-6}$ in $(c).$ \label{fig_ee_random}}
\end{figure}

To demonstrate that the discrete conservation strategy works in more
complicated cases, we consider the thermalization of a random distribution,
of the form:

\begin{equation}
f_{e}^{rand}=\frac{{\mathcal{{P}}}}{\langle\mathcal{{P}},1\rangle_{p}},\qquad\mathrm{{where}}\qquad\mathcal{{P}}(p_{\parallel},p_{\perp})=\frac{J(p_{\parallel},p_{\perp})}{4\pi\Theta K_{2}(1/\Theta)}\exp{\left[-\frac{{\gamma(p)}}{\Theta}\right]},\label{eq:ferandom}
\end{equation}
where $\Theta=1$ and $J$ is a random number function with a uniform
distribution in the range $[0,1]$, see Fig. \ref{fig_ee_random}.
The presence of large gradients in the distribution and small tails
makes this an excellent problem to test discrete conservation errors
and positivity preservation. For a nonlinear relative tolerance of
$10^{-4}$, the relative errors in momentum are larger than in previous
cases, 1 part in $10^{4}$. Tightening the relative nonlinear tolerance
to $10^{-6}$ results in a commensurate decrease of the errors to
1 part in $10^{6}$. 

\subsubsection{Electrical conductivity in weak and strong electric fields}

\begin{figure}
$(a)\hspace{8cm}(b)$

\includegraphics[scale=0.5]{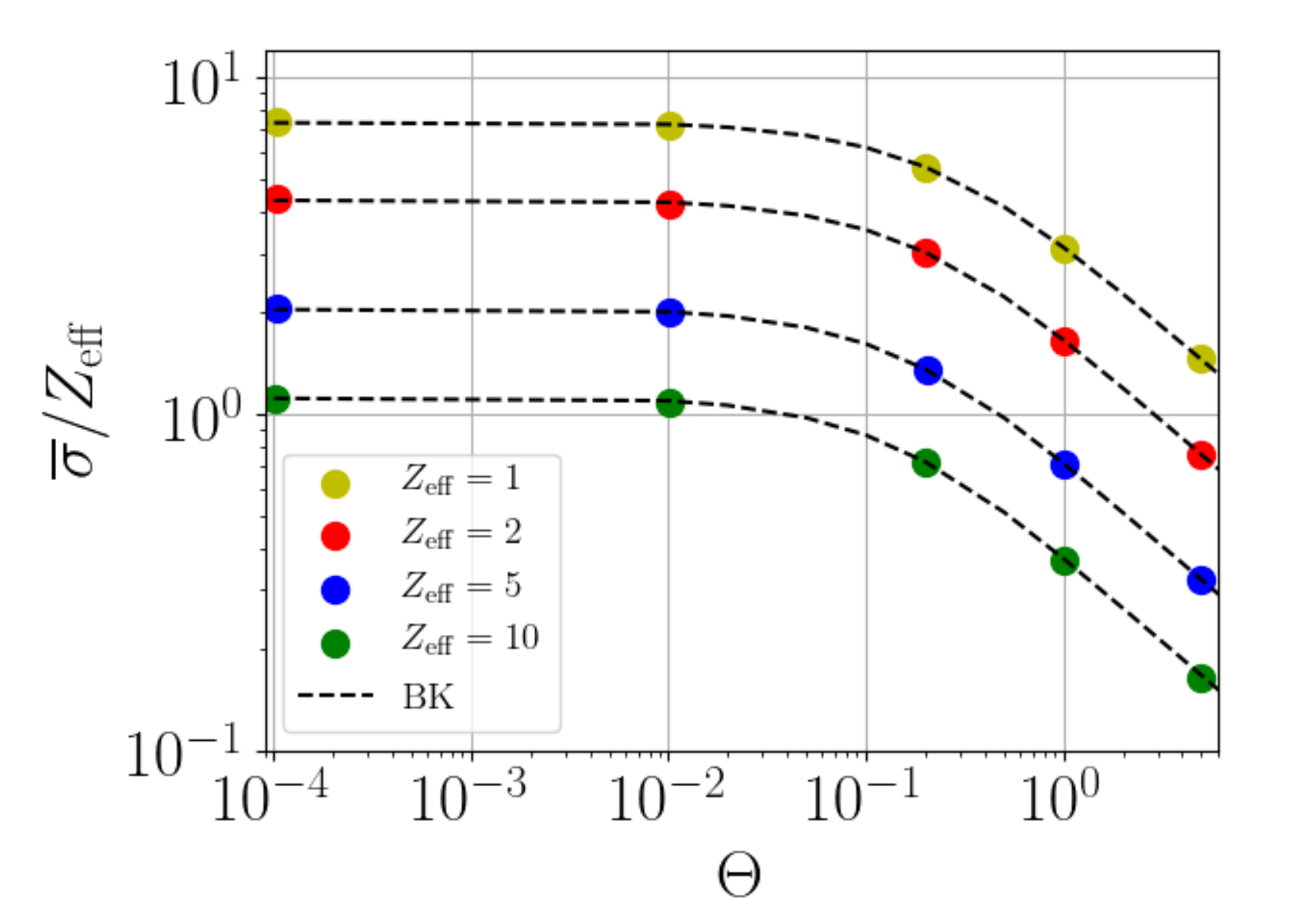}\includegraphics[width=8cm,height=6cm]{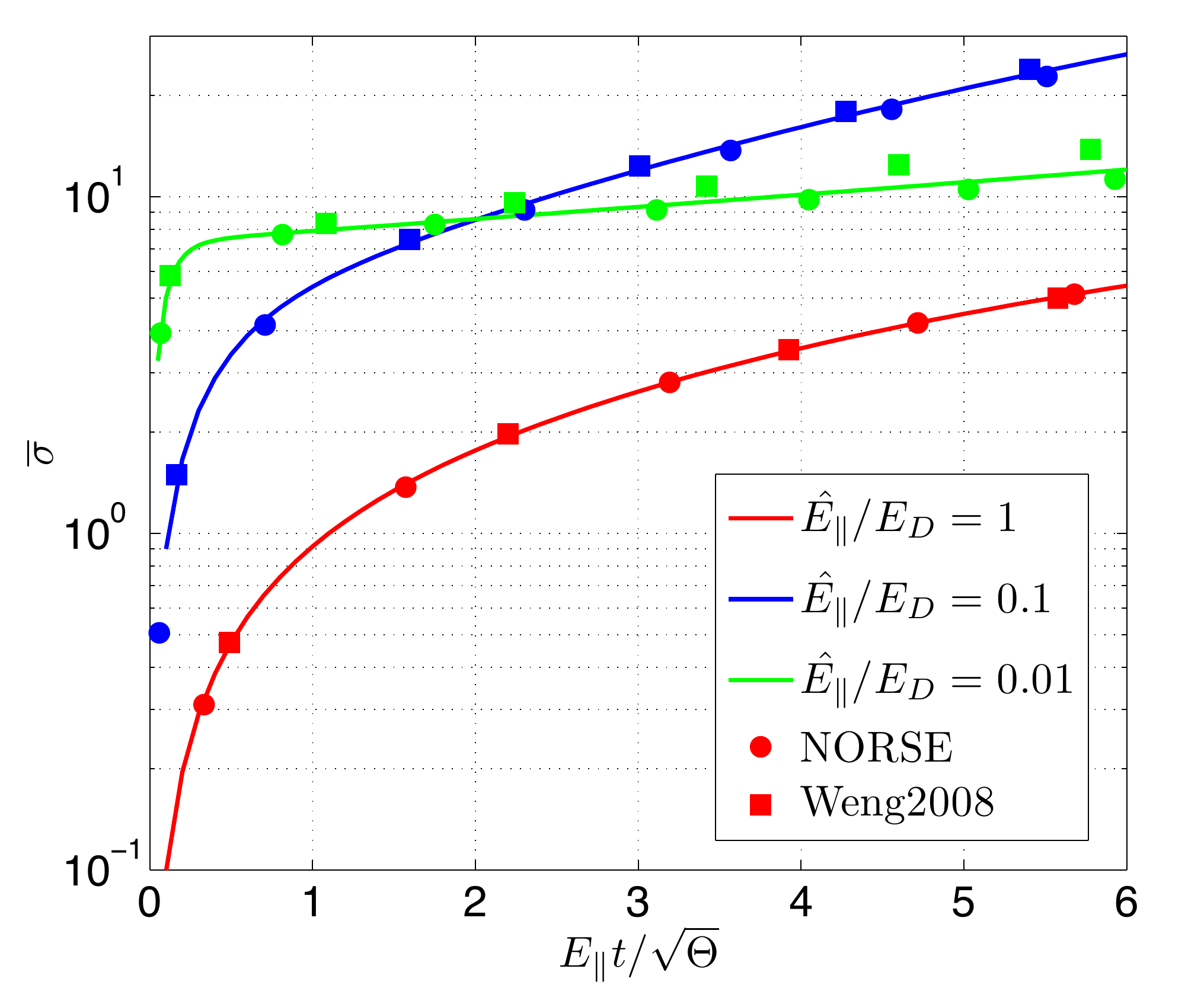}

\caption{Verification under weak and strong electric fields. The computational
domain is uniform with $N_{\parallel}=512$ and $N_{\perp}=256.$
$(a)$ Normalized electrical conductivity vs. $\Theta$ for various
effective ion charges, $Z_{\mathrm{{eff}}}\in[1,10]$, and a weak
electric field of $\hat{{E}}_{\parallel}/E_{D}=0.001$. The momentum
domain sizes vary with a minimum of $p_{\perp}\in(0,0.12)$ and $p_{\parallel}\in(-0.12,0.12)$
for $\Theta=10^{-4}$ and a maximum of $p_{\perp}\in(0,40)$ and $p_{\parallel}\in(-40,40)$
for $\Theta=5$. $(b)$ Time evolution of electrical conductivity
for various strengths of the electric field for an effective ion charge
of $Z_{\mathrm{{eff}}}=1$. The electron distribution was initialized
with a Maxwellian for $\Theta=10^{-4}$ and $n_{e}=1$. The momentum
domains are $p_{\perp}\in(0,0.3)$ and $p_{\parallel}\in(-0.3,0.3)$
for strong electric field ($\hat{{E}}_{\parallel}/E_{D}=1)$, $p_{\perp}\in(0,0.2)$
and $p_{\parallel}\in(-0.2,0.2)$ for intermediate electric field
($\hat{{E}}_{\parallel}/E_{D}=0.1)$, and $p_{\perp}\in(0,0.12)$
and $p_{\parallel}\in(-0.12,0.12)$ for weak electric field ($\hat{{E}}_{\parallel}/E_{D}=0.01)$
.\label{fig_sigma_weak}}
\end{figure}

We consider next the case where collisional friction balances an externally
imposed electric field, leading to finite electrical conductivity.
We verify the code for a wide range of initial temperatures with electrical
conductivity results provided by Braams and Karney \cite{BK89}. To
measure conductivity, we apply a small electric field, $\hat{{E}}_{\parallel}=10^{-3}E_{D}$
where $\hat{{E}}_{\parallel}$ is the parallel electric field, $E_{D}=E_{c}/\Theta$
is the Dreicer field, and $E_{c}$ the Connor-Hastie critical electric
field \cite{CH75}. The electron distribution is initialized using
the Maxwell-Jüttner distribution at various temperatures $\Theta$
and the normalized electrical conductivity is computed as:

\begin{equation}
\bar{{\sigma}}=\frac{{Z_{\mathrm{{eff}}}}}{n_{e}q_{e}\Theta^{3/2}}\frac{\hat{{j}}}{E_{\parallel}},\qquad\qquad\hat{{j}}=-n_{e}q_{e}v_{\parallel},\label{eq:sigma}
\end{equation}
where $v_{\parallel}=p_{\parallel}/\gamma$. A small electric field
deforms the Maxwellian slightly to produce a net electron flow in
the positive $p_{\parallel}$ direction. To prevent numerical overflow,
for $\Theta\apprle10^{-3}$ the initial distribution is defined using
a non-relativistic Maxwellian. Fig. \ref{fig_sigma_weak}$(a)$ depicts
$\bar{\sigma}/Z_{\mathrm{{eff}}}$ at various temperatures. The numerical
simulation results (circles) are in excellent agreement with the analytical
results (lines) from Ref. \cite{BK89}. The electrical conductivity
measurements are made after the simulation reaches a quasi-steady-state
after an initial transient. Because of the applied electric field,
the plasma slowly heats up, and the quasi-steady-state temperatures,
$\Theta=\langle f_{e}(\vec{{p}},t)p^{2}/2\gamma\rangle_{p}$, are
larger (but close) to their initial value. The electrical conductivity,
$\bar{{\sigma}}$, in Eq. (\ref{eq:sigma}) is computed using the
quasi-steady-state temperature.

Figure \ref{fig_sigma_weak}$(b)$ illustrates the time evolution
of electrical conductivity for various electric-field strengths in
the non-relativistic limit. Results of NORSE \cite{NORSE} (circles)
and Weng et al. \cite{Weng} (squares) are also shown. The electron
distribution is initialized with a Maxwellian corresponding to an
initial temperature of $\Theta=10^{-4}$ and $n_{e}=1$. Note that
the Weng et al. study uses the nonrelativistic Fokker-Planck operator.
For all values of electric field, we have good agreement with earlier
studies. For the case of $\hat{{E}}_{\parallel}=0.01E_{D},$ we have
better agreement with NORSE than Weng et al.. Ref. \cite{NORSE} hypothesizes
that the observed deviations between NORSE and Weng et al. in the
small $\hat{{E}}/E_{D}$ limit may be due to numerical heating in
Weng et al.. Our results also suggest the same. 

\subsubsection{Reproducing runaway-electron tail dynamics}

\begin{figure}
\centering{}\includegraphics[scale=0.45]{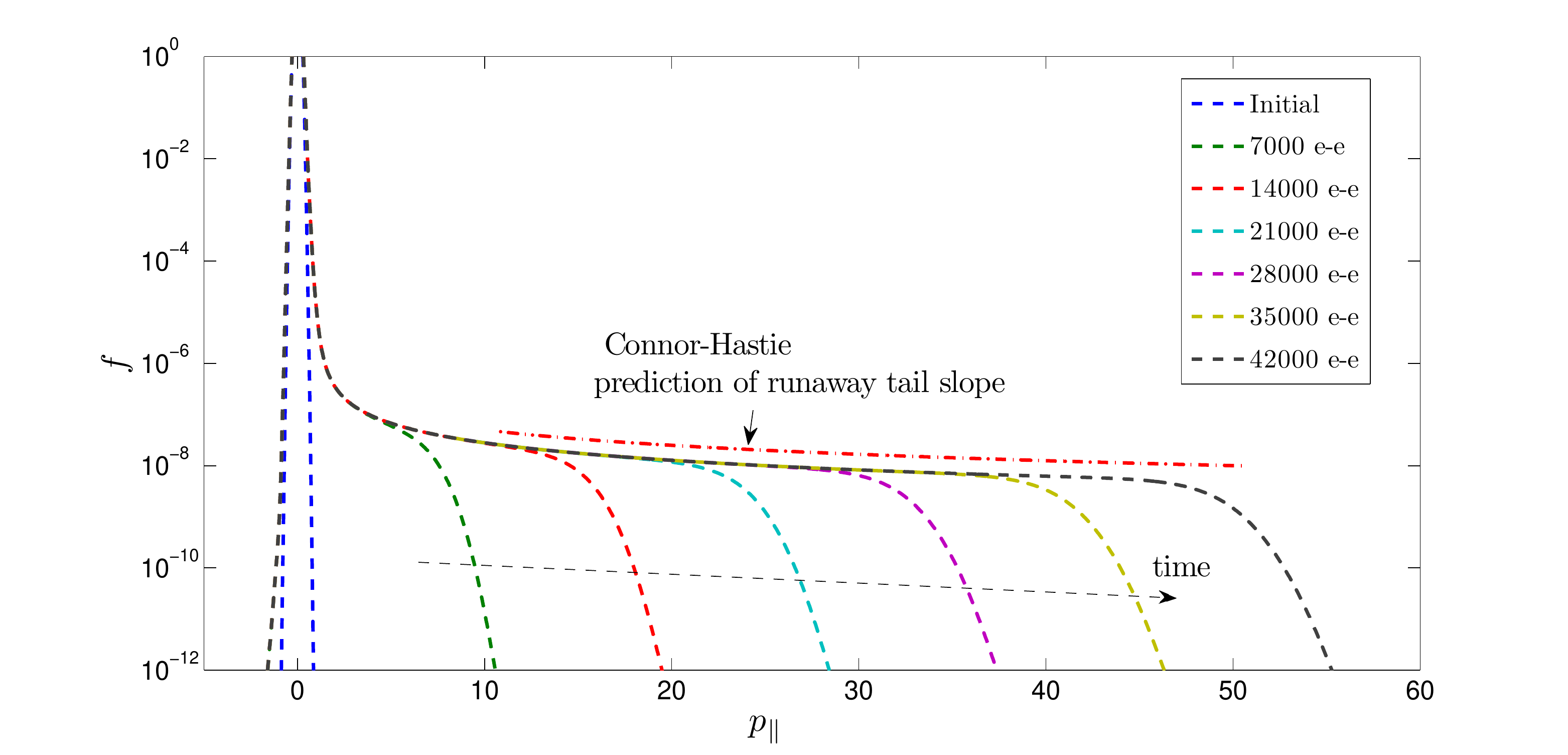}\caption{Verification of runaway tail dynamics. Electron distribution function
vs. $p_{\parallel}$ for $p_{\perp}=\Delta p_{\perp}/2,\,E_{\parallel}=2.25,\,\Theta=0.01$
and $Z_{\mathrm{{eff}}}=1$. The computational domain is uniform with
$N_{\parallel}=2048$ and $N_{\perp}=512$ with $p_{\perp}\in(0,10)$
and $p_{\parallel}\in(-10,60)$. The initial Maxwellian is represented
by the blue line concentrated at the origin, $p_{\parallel}=0.$ At
7000 $\tau_{ee}^{thermal}$ (green line), we see a finite tail develop
from the Maxwellian. This tail grows steadily as time increases. Time
step $\Delta t=0.01$ i.e. 10 $\tau_{ee}^{thermal}$, and a nonlinear
relative convergence tolerance of $r_{NL}=10^{-6}$. \label{fig:runaway_tail}}
\end{figure}

To verify runaway dynamics with existing linear test-particle studies
\cite{CH75,Guo17}, we performed two linearized numerical simulations
where we keep the collisional cofficients fixed in time to those of
a Maxwell-Jüttner distribution with $n_{e}=1$ and $\Theta=0.01$.
In the first simulation, we applied an electric field 2.25 times the
critical value, i.e $E_{\parallel}=2.25.$ This causes some electrons
to overcome the frictional force and accelerate to high speeds. Fig.
\ref{fig:runaway_tail} demonstrates the evolution of the runaway
tail at $42000\,\tau_{ee}^{thermal}$ collision times. The asymptotic
slope of the runaway tail as predicted by Connor-Hastie \cite{CH75}
is:

\[
f_{e}^{tail}\propto\frac{1}{p_{\parallel}}\exp\left(-\frac{{(E_{\parallel}+1)p_{\perp}^{2}}}{2(1+Z_{\mathrm{{eff}}})p_{\parallel}}\right).
\]
As can be seen in the figure, the runaway tail produced by the algorithm
is in excellent agreement with the asymptotic theoretical results.

\begin{figure}
\begin{centering}
\includegraphics[scale=0.47]{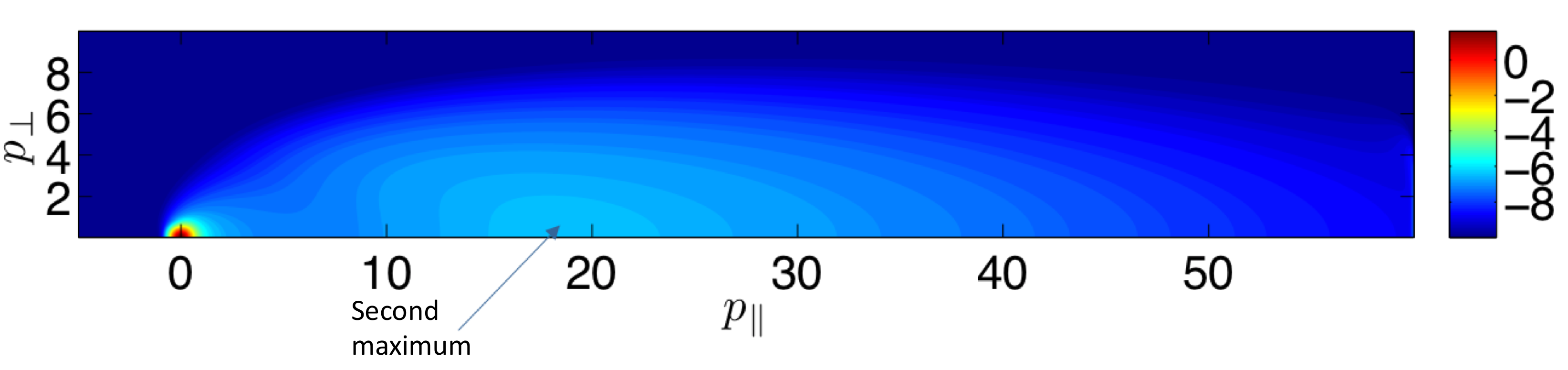}
\par\end{centering}
\caption{Verification of synchrotron radiation physics. Electron distribution
function in the $(p_{\parallel},p_{\perp})$ space for $E_{\parallel}=2.25\,,\Theta=0.01,$
$Z_{\mathrm{{eff}}}=1$ and synchrotron damping factor $S=0.1$. The
second maximum is located at $p\approx18.$ The computational domain
is uniform with $N_{\parallel}=2048$ and $N_{\perp}=512.$ The contour
is plotted at time $t=777.$ Time step, $\Delta t=0.5$ i.e. 500 $\tau_{ee}^{thermal}$,
and a nonlinear relative-convergence tolerance of $r_{NL}=10^{-6}$. }

\label{synchrotron damping}
\end{figure}

In the second simulation, we verify the runaway electron dynamics
in the presence of the synchrotron radiation damping term in Eq. (\ref{eq:synchrotron_damping}).
Because we are interested in the steady-state as $t\rightarrow\infty$,
time marching is performed efficiently with a BDF1 time stepping scheme
(rather than BDF2). Fig. \ref{synchrotron damping} shows the electron
distribution function at $t=777$ (i.e., $\approx777,000\,\tau_{ee}^{thermal}$)
with a damping coefficient of $S=0.1$. Other parameters are $E_{\parallel}=2.25,\Theta=0.01,$
and $Z_{\mathrm{{eff}}}=1$. Ref. \cite{decker2016numerical} performed
a linearized initial value problem and describes the evolution of
runaway electrons in the momentum space as a two-step phenomenon,
beginning with the relative fast formation of a long runaway tail
and a much slower rise of the bump to a steady-state solution. We
find similar behavior here with the electrons accumulating in the
momentum space around $p_{0}\approx18$, to form a second maximum.
The location of the second maximum is in good agreement with Ref.
\cite{Guo17}. Note that, when collisional coefficients are evolved
nonlinearly, we see heating of the bulk (not shown) leading to slide-away
effects as previously reported in Ref. \cite{NORSE}. 

\subsection{Solver performance}

\subsubsection{Algorithmic and parallel scalability.\label{subsec:Algorithmic-and-parallel}}

Table \ref{tab:Parallel-and-algorithmic-random} lists weak parallel
scalability results for the thermalization of a random electron distribution,
see $f_{e}^{rand}$ in Eq. (\ref{eq:ferandom}), with $\Theta=1.$
For simplicity, we employ the BDF1 time-stepping scheme for scalability
tests. We report the wall clock time (WCT) per time step, the number
of nonlinear iterations (NLI) per implicit time step $\Delta t$,
the ratio WCT/NLI (which is an indirect measure of communication costs),
and the ratio between implicit and explicit time steps (which is a
measure of numerical stiffness). The potential linear iterations (PLI)
is the average number of GMRES iterations required for each potential
solve. The absolute tolerances are set to $10^{-8}$. For the initial
solve, the potentials are converged tightly to a relative tolerance
of $10^{-8}$, with a looser tolerance of $10^{-6}$ used for subsequent
nonlinear iterations. The adaptive spline tolerance is defined by
Eq. (\ref{eq:adaptive-spline-tol}) and a fourth-order spline is used
for piece-wise interpolation. The explicit time step is calculated
as:

\[
\Delta t_{\mathrm{{explicit}}}=0.25\min_{\beta=i,e}\left(\frac{{\Delta}p_{\parallel}^{2}}{\max({D_{\beta,\parallel\parallel}})},\frac{{\Delta}p_{\perp}^{2}}{\max({D_{\beta,\perp\perp}})},\frac{{\Delta}p_{\perp}}{\max({A_{\beta,\perp}})},\frac{{\Delta}p_{\parallel}}{\max({A_{\beta,\parallel}})}\right).
\]
Table \ref{tab:Parallel-and-algorithmic-random} demonstrates excellent
parallel scalability (WCT/NLI) up to 4096 processors. When increasing
the number of processors while keeping the problem size per processor
constant, we are effectively increasing the resolution of the problem,
thus making the problem harder to solve for a fixed time step (as
evidenced by the increasing implicit-to-explicit timestep ratio).
This manifests in a very mild growth of the number of nonlinear iterations
(NLI) as we increase the number of cores (the iteration count increases
only by a factor of 3 over a three-order-of-magnitude increase of
the problem size).Table \ref{tab:Parallel-and-algorithmic_MJ} lists
parallel and algorithmic scalability results for the case of collisional
relaxation of two Maxwell-Jüttner distributions, boosted by one momentum
unit in opposite directions and a normalized temperature of $10^{-1}$
in their frames of reference, see $f_{e}^{BMJ}$ in Eq. (\ref{eq:feBMJ}).
We observe good parallel (WCT/NLI) and algorithmic scalability (NLI)
up to 4096 processors with $\Delta t/\Delta t_{\mathrm{{explicit}}}\sim460$
for the high resolution case of $4096\times2048$. 

\begin{figure}
\begin{centering}
\includegraphics[width=8cm,height=6cm]{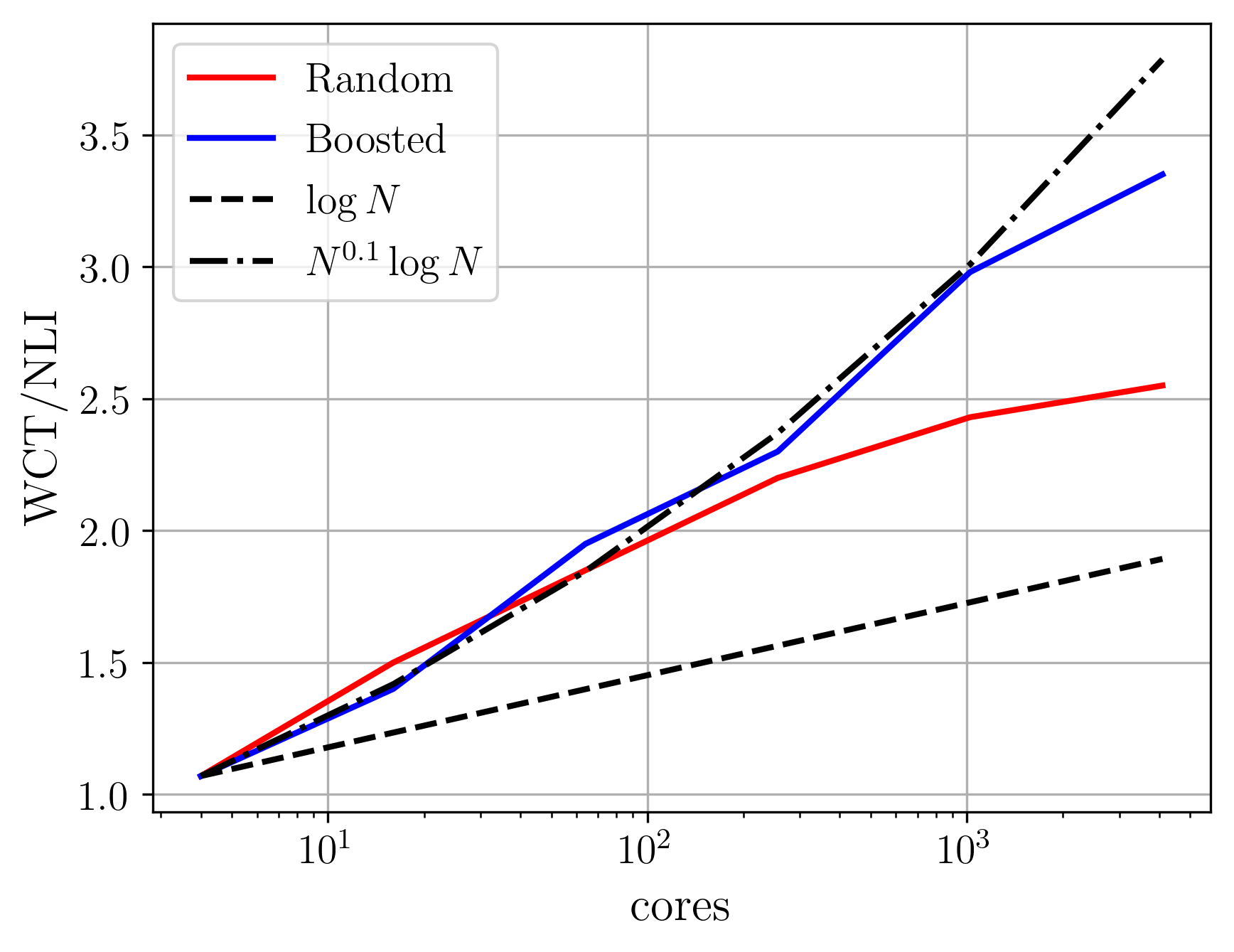}\includegraphics[width=8cm,height=6cm]{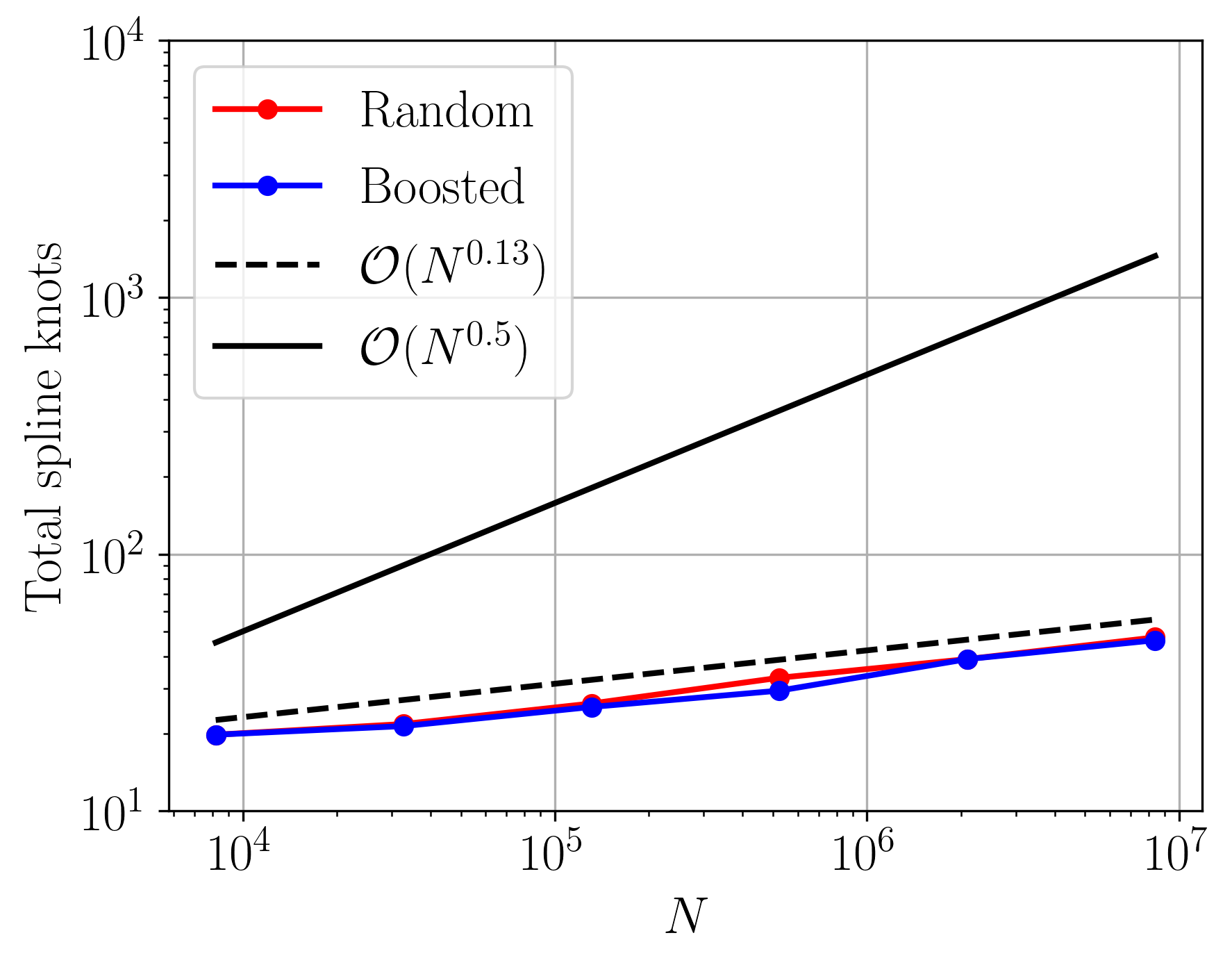}
\par\end{centering}
\caption{Left: Weak parallel scaling test with a domain size per core of $64\times32$
grid points. The figure depicts the wall clock time (WCT) per nonlinear
iteration (NLI) as a function of the number of cores, and demonstrates
that the scaling is consistent with an $\mathcal{O}(N^{0.1}\log N$)
scaling. Right: Cumulative number of adaptive spline knots (for all
potentials) $N_{p}$ vs. the total mesh points, $N,$ demonstrating
an overall scaling of $N_{p}\sim\mathcal{O}(N^{0.13})$.\label{scaling-1}}
\end{figure}

Fig. \ref{scaling-1}$(a)$ illustrates weak scaling results of the
wall clock time per nonlinear iteration (WCT/NLI) vs. the number of
cores for the random electron thermalization (red line, Table \ref{tab:Parallel-and-algorithmic-random})
and the boosted MJ relaxation (blue line, Table \ref{tab:Parallel-and-algorithmic_MJ}).
We observe excellent parallel scalability in both cases. The expected
scaling for parallel multigrid-based solvers is WCT/NLI $\sim\mathcal{O}(\log N)$,
resulting in an overall algorithmic scaling of $\mathcal{{O}}(N\log{N})$.
Instead, we find WCT/NLI $\sim\mathcal{O}(N^{0.1}\log{N})$. The additional
factor of $N^{0.1}$ originates in the growth of the number of spline
knots with $N$, as predicted by Eq. \ref{eq:knot-scaling} and demonstrated
in Fig. \ref{scaling-1}$(b)$. In the figure, we show that the \emph{cumulative}
number of spline knots (for all potentials) increases as $O(N^{0.13})$
for a fourth-order spline, which is more benign than the predicted
one in Eq. \ref{eq:knot-scaling} (which assumed uniformly spaced
knots), and much more efficient than the naive scaling of $\mathcal{O}(N^{3/2})$.
We have also confirmed that the scaling of number of spline knots
with $N$ depends on the spline order, with lower orders resulting
in a larger exponent. For instance, a cubic spline for the same tests
results in $N_{p}$ scaling as $O(N^{0.17})$ (results not shown).
It is interesting to note that, from the figure, the $\mathcal{O}(N^{0.1})$
scaling seems to disappear at large core count (large problem sizes),
which we speculate is due to the problem becoming too well resolved
by the mesh.

\begin{table}
\begin{centering}
\begin{tabular}{|c|c|c|c|c|c|c|c|}
\hline 
$N_{\parallel}$ & $N_{\perp}$ & $np$ & NLI per $\Delta t$ & WCT (sec) per $\Delta t$ & WCT/NLI & $\Delta t/\Delta t_{\mathrm{{explicit}}}$ & PLI\tabularnewline
\hline 
\hline 
128 & 64 & 4 & 4.5 & 4.85 & 1.07 & 82 & 7.8\tabularnewline
\hline 
256 & 128 & 16 & 4 & 6.0 & 1.5 & 329 & 9.14\tabularnewline
\hline 
512 & 256 & 64 & 4 & 7.4 & 1.85 & 1318 & 9.475\tabularnewline
\hline 
1024 & 512 & 256 & 5 & 11.1 & 2.2 & 5270 & 9.088\tabularnewline
\hline 
2048 & 1024 & 1024 & 6.5 & 15.8 & 2.43 & 21083 & 8.969\tabularnewline
\hline 
4096 & 2048 & 4096 & 12 & 30.6 & 2.55 & 84331 & 9.0\tabularnewline
\hline 
\end{tabular}
\par\end{centering}
\caption{Parallel and algorithmic scaling tests: Thermalization of a random
distribution in domain with $p_{\parallel}\in(-10,10)$, $p_{\perp}\in(0,10)$,
and $\Delta t=1.$ The results are averaged over 2 time steps with
nonlinear relative convergence tolerance of $r_{NL}=10^{-5}$ (considering
more time steps is not useful, as the solution has already settled
into a MJ distribution). \label{tab:Parallel-and-algorithmic-random}}

\begin{centering}
\begin{tabular}{|c|c|c|c|c|c|c|c|}
\hline 
$N_{\parallel}$ & $N_{\perp}$ & $np$ & NLI per $\Delta t$ & WCT (sec) per $\Delta t$ & WCT/NLI & $\Delta t/\Delta t_{\mathrm{{explicit}}}$ & PLI\tabularnewline
\hline 
\hline 
128 & 64 & 4 & 3.2 & 3.42 & 1.07 & 0.45 & 8.5\tabularnewline
\hline 
256 & 128 & 16 & 3.3 & 4.63 & 1.4 & 1.8 & 8.6\tabularnewline
\hline 
512 & 256 & 64 & 3 & 5.85 & 1.95 & 7.2 & 10.8\tabularnewline
\hline 
1024 & 512 & 256 & 3 & 6.9 & 2.3 & 29 & 11.6\tabularnewline
\hline 
2048 & 1024 & 1024 & 3 & 8.93 & 2.98 & 115 & 12.4\tabularnewline
\hline 
4096 & 2048 & 4096 & 4 & 13.4 & 3.35 & 460 & 13.55\tabularnewline
\hline 
\end{tabular}
\par\end{centering}
\caption{Parallel and algorithmic scaling tests: Collisional relaxation of
two boosted Maxwell-Jüttner distributions in domain with $\Theta_{b}=0.1$,
$p_{b}=-1,\,1,$ $p_{\parallel}\in(-15,15)$, $p_{\perp}\in(0,15)$,
and $\Delta t=0.01$. Note that the time step chosen is comparable
to the thermal collision time in boosted frame, i.e. $\tau_{ee,b}^{thermal}\approx0.04\tau_{ee}^{relativistic}.$
The results are averaged over 10 time steps with $r_{NL}=10^{-5}$
. \label{tab:Parallel-and-algorithmic_MJ}}
\end{table}

\subsubsection{Spatial and temporal accuracy }

\begin{figure}
$(a)\hspace{8cm}(b)$

\includegraphics[width=8cm,height=6cm]{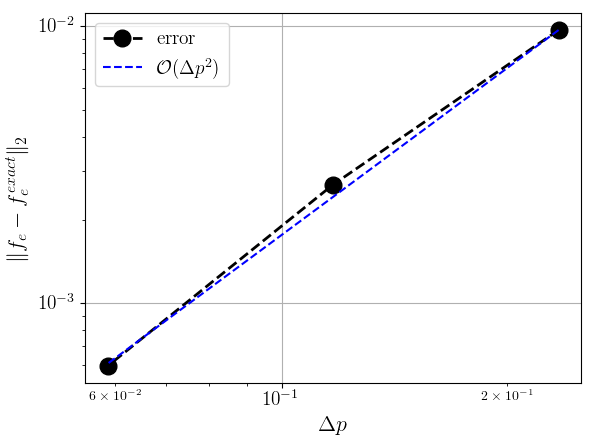}\includegraphics[width=8cm,height=6cm]{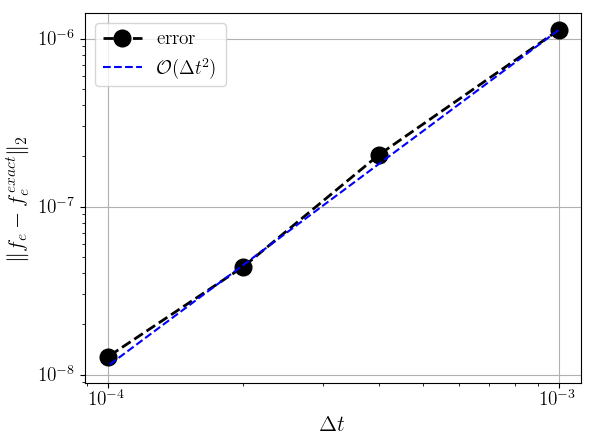}

\caption{Spatial and temporal accuracy measurement of the proposed scheme using
the two boosted MJ configuration. \label{fig:accuracy}}
\end{figure}

Figure \ref{fig:accuracy}$(\emph{{a}}$) illustrates the spatial
accuracy of the proposed scheme with the boosted MJ collisional relaxation
problem described in $\S$\ref{subsec:Algorithmic-and-parallel}.
The `exact' electron distribution $f_{e}^{exact}$ is obtained at
$t=0.15$ (before steady-state is reached) with $N_{\parallel}=1024$,
$N_{\perp}=512$ and $\Delta t=0.01$, while the three data points
(solid dots) correspond to coarser grids of $128\times64$, $256\times128$
and $512\times256$. The blue dashed line corresponds to a second-order
error scaling. The $\ell_{2}-$norm of the error between the `exact'
and numerical electron distributions is computed as: 

\[
\|f_{e}-f_{e}^{exact}\|_{2}=\left(\sum_{j=1}^{N_{\parallel}}\sum_{k=1}^{N_{\perp}}(f_{e,j,k}-f_{e,j,k}^{exact})^{2}2\pi p_{\perp,k}\Delta p_{\parallel,j}\Delta p_{\perp,k}\right)^{0.5}.
\]
We confirm that the proposed implementation is second-order accurate
in space. 

Figure \ref{fig:accuracy}$(\emph{{b}}$) illustrates the temporal
accuracy of the implementation. In this case, the 'exact' electron
distribution $f_{e}^{exact}$ is obtained in a $256\times128$ grid
using the BDF2 time advancement scheme with $\Delta t=5\times10^{-4}$,
see description in $\S$\ref{subsec:Time-stepping-strategy}. The
four data points correspond to larger time steps of $\Delta t=10^{-4},2\times10^{-4},4\times10^{-4},$
and $10^{-3}$. The proposed implementation is confirmed to be second-order
accurate in time.

\section{Conclusions\label{sec:Conclusion-and-Future}}

We have developed a fully implicit, nearly optimal, relativistic nonlinear
Fokker-Plank algorithm with strict conservation properties.. We consider
a 0D2P cylindrical momentum-space representation. The solver employs
the differential form of the Fokker-Planck equation, which requires
the solution of five relativistic potentials in momentum space to
obtain the collisional coefficients. Singularities in the potential
integral formulations are resolved by expressing them in terms of
complete elliptic integrals of the second and third kind. To ensure
a benign scaling of the potential solves with the total number of
mesh points $N$, we employ a multigrid-preconditioned GMRES solver,
and have developed an adaptive spline technique for finding far-field
boundary conditions for the potentials. The adaptive spline technique
results in a small additional exponent in the algorithmic scaling
of $\mathcal{O}(N^{0.1})$. Positivity of the distribution function
is ensured using a continuum-based reformulation approach \cite{DOV}
combined with robust positivity-preserving discretizations schemes
\cite{SMART}. Using an Anderson Acceleration fixed-point iteration
scheme for our nonlinear solves, also preconditioned with multigrid
techniques, we obtain an algorithm that overall scales as $\mathcal{{O}}(N^{1.1}\log{N})$.
The $\log{N}$ contribution is due to the parallel multigrid techniques
employed, and the $N^{0.1}$ contribution is from the proposed adaptive
spline technique. We have demonstrated second-order accuracy in both
space and time, and characterized the performance of our parallel
implementation. We have verified our solver by comparing with previous
results for electrical conductivity measurements in the weak and strong
electric field limits. We have demonstrated the accuracy of conserved
quantities in electron-electron collisions, with small relative errors,
in number density, relativistic momentum, and energy. In addition,
we have examined runaway dynamics and verified it by comparing to
known results \cite{CH75,decker2016numerical,Guo17}. In future work,
we will extend this method to study inhomogeneous plasmas by considering
the spatial dependence of the electron distribution function.

\section*{Acknowledgements}

The authors thank Z. Guo for help in verifying the algorithm and E.
Hirvijoki for insightful discussions on the properties of the relativistic
operator. The authors also thank C. McDevitt and X. Tang for useful
inputs during the course of this project. This work was supported
by the US Department of Energy through the Los Alamos National Laboratory.
Los Alamos National Laboratory is operated by Triad National Security,
LLC, for the National Nuclear Security Administration of U.S. Department
of Energy (Contract No. 89233218CNA000001). This research used resources
provided by the Los Alamos National Laboratory Institutional Computing
Program.

\appendix

\section{Discretization of operators in potential equations and collisional
coefficients}

\subsection*{A1. Linear potential equations. }

The potential operator $L$ consists of Hessian and advective terms,

\[
L\,\psi=({\overline{{I}}}+\vec{{p}}\vec{{p}}):\frac{{\partial^{2}\psi}}{\partial{\vec{{p}}}\partial{\vec{{p}}}}+3\vec{{p}}.\frac{{\partial{\psi}}}{\partial\vec{{p}}}.
\]
The terms are discretized using central differencing. The Laplacian
piece is computed as:

\[
\left(\overline{{\overline{{I}}}}:\frac{{\partial^{2}\psi}}{\partial\vec{{p}}\partial{\vec{{p}}}}\right)_{j,k}=\left(\frac{{\partial^{2}\psi}}{\partial p_{\parallel}^{2}}+\frac{{\partial^{2}\psi}}{\partial p_{\perp}^{2}}\right)_{j,k}=\frac{{X_{j+\frac{{1}}{2},k}-X_{j-\frac{{1}}{2},k}}}{\Delta p{}_{\parallel,j}}+\frac{{p_{\perp,k+\frac{{1}}{2}}Y_{j,k+\frac{{1}}{2}}-p_{\perp,k-\frac{{1}}{2}}Y_{j,k-\frac{{1}}{2}}}}{p_{\perp,k}\Delta p{}_{\perp,k}},
\]
where,

\[
X_{j+\frac{{1}}{2},k}=\frac{{\psi_{j+1,k}-\psi_{j,k}}}{\Delta p_{\parallel,j+\frac{{1}}{2}}},\qquad\qquad Y_{j,k+\frac{{1}}{2}}=\frac{{\left(\psi_{j,k+1}-\psi_{j,k}\right)}}{\Delta p_{\perp,k+\frac{{1}}{2}}}.
\]
The remaining Hessian piece is computed as:

\begin{eqnarray*}
\left(\vec{{p}}\vec{{p}}:\frac{{\partial^{2}\psi}}{\partial\vec{{p}}\partial{\vec{{p}}}}\right)_{j,k} & = & \left[\begin{array}{cc}
p_{\parallel,j}p_{\parallel,j} & p_{\perp,k}p_{\parallel,j}\\
p_{\parallel,j}p_{\perp,k} & p_{\perp,k}p_{\perp,k}
\end{array}\right]:\left[\begin{array}{cc}
\frac{{\partial^{2}\psi}}{\partial p_{\parallel}^{2}} & \frac{{\partial^{2}\psi}}{\partial p_{\parallel}\partial p_{\perp}}\\
\frac{{\partial^{2}\psi}}{\partial p_{\perp}\partial p_{\parallel}} & \frac{{\partial^{2}\psi}}{\partial p_{\perp}^{2}}
\end{array}\right]_{j,k}\\
 & = & p_{\parallel,j}p_{\parallel,j}\frac{{Q_{j+\frac{{1}}{2},k}-Q_{j-\frac{{1}}{2},k}}}{\Delta p{}_{\parallel,j}}+p_{\perp,k}p_{\perp,k}\frac{{R_{j,k+\frac{{1}}{2}}-R_{j,k-\frac{{1}}{2}}}}{\Delta p{}_{\perp,k}}\\
 & + & 2p_{\perp,k}p_{\parallel,j}\frac{{T_{j,k+\frac{{1}}{2}}-T_{j,k-\frac{{1}}{2}}}}{\Delta p{}_{\perp,k}},
\end{eqnarray*}
where

\[
Q_{j+\frac{{1}}{2},k}=\frac{{\psi_{j+1,k}-\psi_{j,k}}}{\Delta p_{\parallel,j+\frac{{1}}{2}}},\quad R_{j,k+\frac{{1}}{2}}=\frac{{\psi_{j,k+1}-\psi_{j,k}}}{\Delta p_{\perp,k+\frac{{1}}{2}}},
\]

\[
T_{j,k+\frac{{1}}{2}}=\frac{{1}}{2}\left(\frac{{\psi_{j+\frac{{1}}{2},k+1}-\psi_{j-\frac{{1}}{2},k+1}}}{\Delta p_{\parallel,j}}+\frac{{\psi_{j+\frac{{1}}{2},k}-\psi_{j-\frac{{1}}{2},k}}}{\Delta p_{\parallel,j}}\right).
\]
The advective piece is computed as:

\[
\left(3\vec{{p}}.\frac{{\partial{\psi}}}{\partial\vec{{p}}}\right)_{j,k}=3\left(p_{\parallel,j}\frac{{\psi_{j+\frac{{1}}{2},k}-\psi_{j-\frac{{1}}{2},k}}}{\Delta p_{\parallel,j}}+p_{\perp,k}\frac{{\psi_{j,k+\frac{{1}}{2}}-\psi_{j,k-\frac{{1}}{2}}}}{\Delta p_{\perp,k}}\right).
\]
Note the cell faced values of $\psi$ are found by linear averaging
across cell centered values, for example $\psi_{j+1/2,k}=0.5(\psi_{j,k}+\psi_{j+1,k})$
and $\psi_{j,k+1/2}=0.5(\psi_{j,k}+\psi_{j,k+1})$. 

\subsection*{A2. Collisional coefficients. }

Once the potentials are determined, the friction coefficients are
evaluated using Eq. (\ref{eq:A-1}). The components of $\vec{{K}}\psi$
at the cell center are defined as:

\begin{eqnarray*}
(K\psi)_{j,k} & = & \left(\overline{{\overline{{I}}}}+\vec{{p}}\vec{{p}}\right)_{j,k}\cdot\left(\frac{{\partial\psi}}{\partial\vec{{p}}}\right)_{j,k}\\
 & = & \left[\begin{array}{c}
(1+p_{\parallel,j}p_{\parallel,j})\left(\frac{{\psi_{j+\frac{{1}}{2},k}-\psi_{j-\frac{{1}}{2},k}}}{\Delta p_{\parallel,j}}\right)+p_{\parallel,j}p_{\perp,k}\left(\frac{{\psi_{j,k+\frac{{1}}{2}}-\psi_{j,k-\frac{{1}}{2}}}}{\Delta p_{\perp,k}}\right)\\
(1+p_{\perp,k}p_{\perp,k})\left(\frac{{\psi_{j,k+\frac{{1}}{2}}-\psi_{j,k-\frac{{1}}{2}}}}{\Delta p_{\perp,k}}\right)+p_{\parallel,j}p_{\perp,k}\left(\frac{{\psi_{j+\frac{{1}}{2},k}-\psi_{j-\frac{{1}}{2},k}}}{\Delta p_{\parallel,j}}\right)
\end{array}\right]
\end{eqnarray*}
The cell face values of $\psi$ are found by taking the average of
cell-centered values. A similar discretization approach is used when
evaluating the diffusion coefficient, Eq. (\ref{eq:D-1}) .

\subsection*{A3. Reformulated off-diagonal tensor diffusion terms (effective friction
coefficients). }

The off-diagonal diffusion coefficients are expressed as effective
friction coefficients of the form $D_{\parallel\perp}\partial\ln{f}/\partial p_{\perp}$
and $D_{\perp\parallel}\partial\ln{f}/\partial p_{\parallel}$, see
Eq (\ref{eq:Feff}). The momentum-space derivatives of $\ln{f}$ at
cell centers are evaluated by averaging the cell vertex values, for
example:

\[
\left(\frac{{\partial\ln{f}}}{\partial p_{\perp}}\right)_{j,k}=\frac{{1}}{4}\left(\left(\frac{{\partial\ln{f}}}{\partial p_{\perp}}\right)_{j+\frac{{1}}{2},k+\frac{{1}}{2}}+\left(\frac{{\partial\ln{f}}}{\partial p_{\perp}}\right)_{j-\frac{{1}}{2},k+\frac{{1}}{2}}+\left(\frac{{\partial\ln{f}}}{\partial p_{\perp}}\right)_{j-\frac{{1}}{2},k-\frac{{1}}{2}}+\left(\frac{{\partial\ln{f}}}{\partial p_{\perp}}\right)_{j+\frac{{1}}{2},k-\frac{{1}}{2}}\right),
\]
where the cell vertex value is obtained by averaging over adjacent
face-centered values:

\[
\left(\frac{{\partial\ln{f}}}{\partial p_{\perp}}\right)_{j+\frac{{1}}{2},k+\frac{{1}}{2}}=\frac{{1}}{2}\left(\frac{{\ln(\lvert{f_{j,k+1}}\rvert+\epsilon_{l})-\ln{(\lvert f_{j,k}\rvert+\epsilon_{l})}}}{\Delta p_{\perp,k+\frac{{1}}{2}}}+\frac{{\ln{(\vert{f}}_{j+1,k+1}\rvert+\epsilon_{l})-\ln{(\lvert{f}_{j+1,k}}\rvert+\epsilon_{l})}}{\Delta p_{\perp,k+\frac{{1}}{2}}}\right),
\]

\[
\left(\frac{{\partial\ln{f}}}{\partial p_{\perp}}\right)_{j+\frac{{1}}{2},k-\frac{{1}}{2}}=\frac{{1}}{2}\left(\frac{{\ln(\vert{f}_{j,k}\rvert+\epsilon_{l})-\ln{(\lvert f_{j,k-1}\rvert+\epsilon_{l})}}}{\Delta p_{\perp,k-\frac{{1}}{2}}}+\frac{{\ln{(\vert{f}}_{j+1,k}\rvert+\epsilon_{l})-\ln{(\vert{f}_{j+1,k-1}}\rvert+\epsilon_{l})}}{\Delta p_{\perp,k-\frac{{1}}{2}}}\right),
\]

\[
\left(\frac{{\partial\ln{f}}}{\partial p_{\perp}}\right)_{j-\frac{{1}}{2},k+\frac{{1}}{2}}=\frac{{1}}{2}\left(\frac{{\ln(\vert{f}_{j-1,k+1}\rvert+\epsilon_{l})-\ln{(\lvert f_{j-1,k}\rvert+\epsilon_{l})}}}{\Delta p_{\perp,k+\frac{{1}}{2}}}+\frac{{\ln{(\lvert{f}}_{j,k+1}\rvert+\epsilon_{l})-\ln{(\vert{f}_{j,k}}\rvert+\epsilon_{l})}}{\Delta p_{\perp,k+\frac{{1}}{2}}}\right),
\]

\[
\left(\frac{{\partial\ln{f}}}{\partial p_{\perp}}\right)_{j-\frac{{1}}{2},k-\frac{{1}}{2}}=\frac{{1}}{2}\left(\frac{{\ln(\vert{f_{j-1,k}}\rvert+\epsilon_{l})-\ln{(\vert f_{j-1,k-1}\rvert+\epsilon_{l})}}}{\Delta p_{\perp,k-\frac{{1}}{2}}}+\frac{{\ln{(\lvert{f}}_{j,k}\rvert+\epsilon_{l})-\ln{(\vert{f}_{j,k-1}}\rvert+\epsilon_{l})}}{\Delta p_{\perp,k-\frac{{1}}{2}}}\right),
\]
where $\epsilon_{l}=10^{-30}$ is added to mollify singularities.
Once computed at the cell centers, the friction coefficients at the
cell faces are found by linear averaging. 

\section{Solution of singular integrals in relativistic potentials}

\label{app:potential_elliptic_kernel}

\subsection*{B1. Solution of first singular integral}

We seek a solution of the integral:
\begin{equation}
I=\int_{0}^{2\pi}\frac{d\phi}{\sqrt{r^{2}-1}},\label{eq:integral}
\end{equation}
with:
\[
r=\sqrt{(1+p^{2})(1+(p')^{2})}-\mathbf{p}\cdot\mathbf{p}'=\underbrace{\sqrt{(1+p^{2})(1+(p')^{2})}-p_{\parallel}p'_{\parallel}}_{a^{2}}-\underbrace{p_{\perp}p'_{\perp}}_{b^{2}}\cos\phi=a^{2}-b^{2}\cos\Phi.
\]
We consider the case of $b^{2}>0$. Note $r^{2}-1=(r+1)(r-1)$. Since
$r\geq1$, it follows that:
\begin{equation}
a^{2}\geq b^{2}+1.\label{eq:coeffs_constraint}
\end{equation}

To begin, we consider the change of variable $t=\cos\phi$. We consider
the following cases:
\begin{eqnarray*}
t=\cos\phi & , & \phi\in[0,\frac{\pi}{2}],\phi\in[\frac{3\pi}{2},2\pi]\,;\,d\phi=\frac{-dt}{\sqrt{1-t^{2}}}\\
t=-\cos\phi & , & \phi\in[\frac{\pi}{2},\frac{3\pi}{2}]\,;\,d\phi=\frac{dt}{\sqrt{1-t^{2}}}.
\end{eqnarray*}
This gives:
\[
I=2\underbrace{\int_{0}^{1}\frac{dt}{\sqrt{(1-t^{2})(a^{2}+1-b^{2}t)(a^{2}-1-b^{2}t)}}}_{I_{1}}+2\underbrace{\int_{0}^{1}\frac{dt}{\sqrt{(1-t^{2})(a^{2}+1+b^{2}t)(a^{2}-1+b^{2}t)}}}_{I_{2}}.
\]

\subsubsection*{Solution of $I_{2}$ integral}

We begin with the integral $I_{2}$. We follow Abramowitz \& Stegun
\cite{abramowitz}, and consider the polynomials:
\begin{eqnarray*}
Q_{1} & = & 1-t^{2},\\
Q_{2} & = & (a^{2}+1+b^{2}t)(a^{2}-1+b^{2}t).
\end{eqnarray*}
These polynomials have real roots $\pm1$, $-\frac{a^{2}+1}{b^{2}}$,
$-\frac{a^{2}-1}{b^{2}}$. Because of Eq. (\ref{eq:coeffs_constraint}),
it is apparent that the last two roots are $\leq-1$, and hence $Q_{1}$
and $Q_{2}$ do not have nested roots. In this case, one can consider
the transformation to the canonical forms of the elliptic integrals
by constructing the polynomial:
\begin{equation}
Q_{1}-\lambda Q_{2}=-(1+\lambda b^{4})t^{2}-2b^{2}a^{2}\lambda t+1-\lambda(a^{4}-1).\label{eq:Q1-Q2-transformation}
\end{equation}
Seeking a zero discriminant for the quadratic form in $t$ gives the
following value for $\lambda$:
\begin{eqnarray}
b^{4}a^{4}\lambda^{2} & = & (\lambda(a^{4}-1)-1)(1+\lambda b^{4})\Rightarrow\lambda^{2}b^{4}-\lambda(a^{4}-b^{4}-1)+1=0\nonumber \\
 & \Rightarrow & \lambda_{\pm}=\frac{(a^{4}-b^{4}-1)\pm\sqrt{(a^{4}-b^{4}-1)^{2}-4b^{4}}}{2b^{4}}.\label{eq:lambda-def}
\end{eqnarray}
Note that these roots are real and semi-positive, since, by Eq. (\ref{eq:coeffs_constraint}):
\[
(a^{4}-b^{4}-1)\geq2b^{2}.
\]
Also, it is clear that 
\begin{equation}
\lambda_{+}>\lambda_{-}>0,\label{eq:lambda-root-ranking}
\end{equation}
and that:
\begin{equation}
\lambda_{+}\lambda_{-}=\frac{1}{b^{4}}.\label{eq:lambda-root-product}
\end{equation}
Since the discriminant for Eq. (\ref{eq:Q1-Q2-transformation}) vanishes,
it follows that the roots of $Q_{1}-\lambda Q_{2}$ are perfect squares
and are given by:
\begin{equation}
t=-t_{\pm}\,\,;\,\,t_{\pm}=\frac{\lambda_{\pm}b^{2}a^{2}}{1+\lambda_{\pm}b^{4}}.\label{eq:t-roots}
\end{equation}
Therefore:
\begin{eqnarray}
Q_{1}-\lambda_{+}Q_{2} & = & -(1+\lambda b^{4})(t+t_{+})^{2},\label{eq:Q1-Q2-plus}\\
Q_{1}-\lambda_{-}Q_{2} & = & -(1+\lambda b^{4})(t+t_{-})^{2}.\label{eq:Q1-Q2-minus}
\end{eqnarray}

At this point, it is useful to point out a few properties of the roots
$t_{\pm}$ in Eq. (\ref{eq:t-roots}). Firstly, from Eq. (\ref{eq:lambda-root-ranking})
it follows that:
\begin{equation}
t_{+}>t_{-}.\label{eq:t-root-ranking}
\end{equation}
Secondly, from the polynomial in Eq. (\ref{eq:Q1-Q2-transformation})
and the properties of the quadratic equations, we can write:
\begin{equation}
t_{\pm}^{2}=\frac{\lambda_{\pm}(a^{4}-1)-1}{1+\lambda_{\pm}b^{4}},\label{eq:t-squared}
\end{equation}
which can be used to prove that: 
\begin{equation}
t_{-}\leq1\label{eq:tminus<1}
\end{equation}
(needed for later) as follows:
\begin{equation}
t_{-}^{2}\leq1\Leftrightarrow\lambda_{-}\underbrace{(a^{4}-b^{4}-1)}_{\geq2b^{2}}<2\Leftrightarrow\lambda_{-}\leq1/b^{2},\label{eq:lambda_minus_ineq}
\end{equation}
which can be shown to be true when noting that:
\[
(a^{4}-b^{4}-1)^{2}-4b^{4}=(a^{4}-b^{4}-1-2b^{2})(a^{4}-b^{4}-1+2b^{2})\geq(a^{4}-b^{4}-1-2b^{2})^{2}.
\]
The inequality follows from Eq. (\ref{eq:lambda-def}). Finally, using
Eq. (\ref{eq:lambda_minus_ineq}) and the inequality above, we can
also readily prove that:
\begin{equation}
t_{+}=\frac{\lambda_{+}b^{2}a^{2}}{1+\lambda_{+}b^{4}}=\frac{a^{2}}{b^{2}}\frac{1}{\lambda_{-}+1}\geq\frac{a^{2}}{b^{2}+1}\geq1,\label{eq:tplus>one}
\end{equation}
which will be important later.

Eqs (\ref{eq:Q1-Q2-plus}, \ref{eq:Q1-Q2-minus}) can be solved for
$Q_{1}$ and $Q_{2}$ as follows:
\begin{eqnarray*}
Q_{2} & = & a_{2+}(t+t_{+})^{2}-a_{2-}(t+t_{-})^{2},\\
Q_{1} & = & a_{1+}(t+t_{+})^{2}-a_{1-}(t+t_{-})^{2}.
\end{eqnarray*}
Here:
\[
a_{2\pm}=\frac{1+\lambda_{\pm}b^{4}}{\lambda_{+}-\lambda_{-}}\,\,;\,\,a_{1\pm}=\frac{\lambda_{\mp}(1+\lambda_{\pm}b^{4})}{\lambda_{+}-\lambda_{-}}.
\]
Note that:
\begin{itemize}
\item $a_{2+}>a_{2-}$ (from Eq. \ref{eq:lambda-root-ranking}).
\item $a_{1+}/a_{1-}=t_{-}/t_{+}<1$ (from Eq. \ref{eq:t-root-ranking}).
\end{itemize}
From the expressions of $Q_{1}$, $Q_{2}$, one can write:
\[
Q_{1}Q_{2}=(t+t_{+})^{4}\left[a_{2+}-a_{2-}\frac{(t+t_{-})^{2}}{(t+t_{+})^{2}}\right]\left[a_{1+}-a_{1-}\frac{(t+t_{-})^{2}}{(t+t_{+})^{2}}\right].
\]
Following Ref. \cite{abramowitz}, we postulate the change of variables:
\[
w=\frac{(t+t_{-})}{(t+t_{+})}\Rightarrow dw=\frac{t_{+}-t_{-}}{(t_{+}+t)^{2}}dt.
\]
Hence:
\begin{equation}
I_{2}=\int_{0}^{1}\frac{dt}{\sqrt{Q_{1}Q_{2}}}=\frac{1}{t_{+}-t_{-}}\int_{w_{0}}^{w_{1}}\frac{dw}{\sqrt{\left[a_{2+}-a_{2-}w^{2}\right]\left[a_{1+}-a_{1-}w^{2}\right]}}.\label{eq:I2-dw}
\end{equation}
Here:
\[
w_{0}=\frac{t_{-}}{t_{+}}<1\,\,;\,\,w_{1}=\frac{1+t_{-}}{1+t_{+}}\,\,;\,\,w_{0}<w_{1}<1.
\]
The result in Eq. (\ref{eq:I2-dw}) can be written as a canonical
elliptic integral by considering:
\begin{equation}
\frac{a_{1+}}{a_{1-}}=\frac{t_{-}}{t_{+}}=w_{0}=e^{2}<1\,\,;\,\,\frac{a_{2+}}{a_{2-}}=\frac{1+\lambda_{+}b^{4}}{1+\lambda_{-}b^{4}}=d^{2}>1,\label{eq:defs-e-d}
\end{equation}
to find:
\[
I_{2}=\int_{0}^{1}\frac{dt}{\sqrt{Q_{1}Q_{2}}}=\frac{1}{(t_{+}-t_{-})\sqrt{a_{2-}a_{1-}}}\int_{e^{2}}^{e}\frac{dw}{\sqrt{\left[d^{2}-w^{2}\right]\left[e^{2}-w^{2}\right]}}.
\]
Here, we have used the surprising property that:
\[
w_{1}^{2}=\left(\frac{1+t_{-}}{1+t_{+}}\right)^{2}=\frac{t_{-}}{t_{+}}=e^{2}\Rightarrow w_{1}=e,
\]
which can be demonstrated by using the definition of $t_{\pm}$ (Eq.
\ref{eq:t-roots}) and $t_{\pm}^{2}$ (Eq. \ref{eq:t-squared}). 

\subsubsection*{Solution of $I_{1}$ integral}

The solution of the integral $I_{1}$ follows a similar development,
except now:
\begin{eqnarray*}
Q_{1} & = & 1-t^{2},\\
Q_{2} & = & (a^{2}+1-b^{2}t)(a^{2}-1-b^{2}t).
\end{eqnarray*}
With these definitions, it can be shown that the discriminant of the
combination $Q_{1}-\lambda Q_{2}$ is exactly the same, and therefore
so are the solutions $\lambda_{\pm}$. However the roots in $t$ now
have opposite signs:
\begin{equation}
t=t_{\pm}\,\,;\,\,t_{\pm}=\frac{\lambda_{\pm}b^{2}a^{2}}{1+\lambda_{\pm}b^{4}},\label{eq:t-roots-1}
\end{equation}
and the factorization of $Q_{1,2}$ reads:
\begin{eqnarray*}
Q_{2} & = & a_{2+}(t-t_{+})^{2}-a_{2-}(t-t_{-})^{2},\\
Q_{1} & = & a_{1+}(t-t_{+})^{2}-a_{1-}(t-t_{-})^{2}.
\end{eqnarray*}
From the expressions of $Q_{1}$, $Q_{2}$, one can write:
\[
Q_{1}Q_{2}=(t-t_{+})^{4}\left[a_{2+}-a_{2-}\frac{(t-t_{-})^{2}}{(t-t_{+})^{2}}\right]\left[a_{1+}-a_{1-}\frac{(t-t_{-})^{2}}{(t-t_{+})^{2}}\right].
\]
Following Ref. \cite{abramowitz}, we postulate the change of variables:
\[
w=\frac{(t-t_{-})}{(t_{+}-t)}\Rightarrow dw=\frac{t_{+}-t_{-}}{(t_{+}-t)^{2}}dt.
\]
When postulating this change of variables, we have taken into account
the fact that $t\leq1<t_{+}$ (Eq. \ref{eq:tplus>one}), and that
$t_{-}<1$ (Eq. \ref{eq:tminus<1}). It follows that:
\begin{equation}
I_{1}=\int_{0}^{1}\frac{dt}{\sqrt{Q_{1}Q_{2}}}=\frac{1}{t_{+}-t_{-}}\int_{w_{0}}^{w_{1}}\frac{dw}{\sqrt{\left[a_{2+}-a_{2-}w^{2}\right]\left[a_{1+}-a_{1-}w^{2}\right]}},\label{eq:I1-dw}
\end{equation}
where:
\[
w_{0}=-\frac{t_{-}}{t_{+}}=-e^{2}<0\,\,;\,\,w_{1}=\frac{1-t_{-}}{t_{+}-1}>0.
\]
As before, one can readily prove that:
\[
w_{1}^{2}=\left(\frac{1-t_{-}}{t_{+}-1}\right)^{2}=\frac{t_{-}}{t_{+}}=e^{2},
\]
and therefore $w_{1}=e$. There results:
\begin{equation}
I_{1}=\int_{0}^{1}\frac{dt}{\sqrt{Q_{1}Q_{2}}}=\frac{1}{(t_{+}-t_{-})\sqrt{a_{2-}a_{1-}}}\int_{-e^{2}}^{e}\frac{dw}{\sqrt{\left[d^{2}-w^{2}\right]\left[e^{2}-w^{2}\right]}}.\label{eq:I1-dw-1}
\end{equation}

\subsubsection*{Solution of total integral $I$}

When combining these two solutions, we find:
\begin{eqnarray*}
I & = & 2(I_{1}+I_{2})=\frac{2}{(t_{+}-t_{-})\sqrt{a_{2-}a_{1-}}}\left[\int_{e^{2}}^{e}\frac{dw}{\sqrt{\left[d^{2}-w^{2}\right]\left[e^{2}-w^{2}\right]}}+\int_{-e^{2}}^{e}\frac{dw}{\sqrt{\left[d^{2}-w^{2}\right]\left[e^{2}-w^{2}\right]}}\right]\\
 & = & \frac{2}{(t_{+}-t_{-})\sqrt{a_{2-}a_{1-}}}\left[\int_{e^{2}}^{e}+\int_{0}^{e}+\int_{-e^{2}}^{0}\right]=\frac{2}{(t_{+}-t_{-})\sqrt{a_{2-}a_{1-}}}\left[\int_{e^{2}}^{e}+\int_{0}^{e}+\int_{0}^{e^{2}}\right]\\
 & = & \frac{4}{(t_{+}-t_{-})\sqrt{a_{2-}a_{1-}}}\int_{0}^{e}\frac{dw}{\sqrt{\left[d^{2}-w^{2}\right]\left[e^{2}-w^{2}\right]}},
\end{eqnarray*}
which can be written in terms of the complete elliptic integral of
the second kind as \cite{abramowitz}:
\[
I=\frac{4K(m)}{d(t_{+}-t_{-})\sqrt{a_{2-}a_{1-}}}=\frac{4K(m)}{(t_{+}-t_{-})\sqrt{a_{2+}a_{1-}}},
\]
where in the last step we have used the definition of $d$ (Eq. \ref{eq:defs-e-d}),
and where:
\[
m=e^{2}/d^{2}.
\]

\subsection*{B2. Solution of second singular integral}

In the previous section, we determined the root structure of the radicand
and removed the odd terms in the radicand. We employ this approach
and also use ideas from Ref. \cite{labahn-1} to express the following
elliptic integral, 

\[
H=\int_{o}^{2\pi}\frac{{r\,d\phi}}{\sqrt{{r^{2}-1}}},
\]
 in terms of Legendre's elliptic functions. Recall that

\[
r=a^{2}-b^{2}\cos{\phi}.
\]
The integral can thus be expressed as,

\begin{eqnarray*}
H & = & 2\int_{0}^{1}\frac{(a^{2}-b^{2}t)dt}{\sqrt{(1-t^{2})(a^{2}+1-b^{2}t)(a^{2}-1-b^{2}t)}}+2\int_{0}^{1}\frac{(a^{2}+b^{2}t)dt}{\sqrt{(1-t^{2})(a^{2}+1+b^{2}t)(a^{2}-1+b^{2}t)}}.
\end{eqnarray*}
We can regroup as we know from $\S$B1 the solution when the numerator
is unity, 

\begin{equation}
H=a^{2}I+2\underbrace{\int_{0}^{1}\frac{-b^{2}t\,dt}{\sqrt{(1-t^{2})(a^{2}+1-b^{2}t)(a^{2}-1-b^{2}t)}}}_{H_{1}}+2\underbrace{\int_{0}^{1}\frac{b^{2}t\,dt}{\sqrt{(1-t^{2})(a^{2}+1+b^{2}t)(a^{2}-1+b^{2}t)}}}_{H_{2}}.\label{eq:H}
\end{equation}
Removing the odd terms in the radicand, we obtain, 

\[
H_{1}=\int_{0}^{1}\frac{-b^{2}t\,dt}{\sqrt{Q_{1}Q_{2}}}=\frac{-b^{2}}{(t_{+}-t_{-})\sqrt{a_{2-}a_{1-}}}\int_{-e^{2}}^{e}\frac{(wt_{+}+t_{-})/(1+w)dw}{\sqrt{\left[d^{2}-w^{2}\right]\left[e^{2}-w^{2}\right]}}
\]

\[
=\frac{-b^{2}}{(t_{+}-t_{-})\sqrt{a_{2-}a_{1-}}}\int_{-e^{2}}^{e}\frac{R_{1}(w)dw}{\sqrt{\left[d^{2}-w^{2}\right]\left[e^{2}-w^{2}\right]}},
\]
and

\[
H_{2}=\int_{0}^{1}\frac{b^{2}t\,dt}{\sqrt{Q_{1}Q_{2}}}=\frac{b^{2}}{(t_{+}-t_{-})\sqrt{a_{2-}a_{1-}}}\int_{e^{2}}^{e}\frac{(-wt_{+}+t_{-})/(-1+w)dw}{\sqrt{\left[d^{2}-w^{2}\right]\left[e^{2}-w^{2}\right]}}
\]

\[
=\frac{b^{2}}{(t_{+}-t_{-})\sqrt{a_{2-}a_{1-}}}\int_{e^{2}}^{e}\frac{R_{2}(w)dw}{\sqrt{\left[d^{2}-w^{2}\right]\left[e^{2}-w^{2}\right]}}.
\]
The rational functions of $w$, $R_{1}$ and $R_{2}$, can be expressed
in terms of odd and even functions. This is because the odd term can
be simplified into elementary functions via trigonometric substitutions,
see Ref. \cite{labahn-1}. 

\[
R_{1}(w)=\frac{w}{1-w^{2}}(t_{+}-t_{-})+\frac{{t_{-}-w^{2}t_{+}}}{1-w^{2}}
\]

\[
R_{2}(w)=\frac{w}{1-w^{2}}(t_{+}-t_{-})-\frac{{t_{-}-w^{2}t_{+}}}{1-w^{2}}
\]
However, in our case we observe that these terms cancel each other.
Examining the odd terms in $R_{1}$ and $R_{2}$, and adding together
their contribution to $H$, we find:

\[
H_{1}^{odd}+H_{2}^{odd}=-\int_{-e^{2}}^{e}+\int_{e^{2}}^{e}=-\int_{-e^{2}}^{e^{2}}=0,
\]
as the odd function is asymmetric about the origin. The even term
can be further factorized into:

\[
\frac{t_{-}-w^{2}t_{+}}{1-w^{2}}=t_{+}-\frac{t_{+}-t_{-}}{1-w^{2}}.
\]
Putting the above expression into $H_{1}$ and $H_{2}$ , the contributions
of the even terms may be expressed as, 

\[
H_{1}+H_{2}=-\frac{b^{2}t_{+}}{(t_{+}-t_{-})\sqrt{a_{2-}a_{1-}}}\left(\int_{-e^{2}}^{e}\frac{dw}{\sqrt{\left[d^{2}-w^{2}\right]\left[e^{2}-w^{2}\right]}}+\int_{e^{2}}^{e}\frac{dw}{\sqrt{\left[d^{2}-w^{2}\right]\left[e^{2}-w^{2}\right]}}\right)
\]

\[
+\frac{b^{2}}{\sqrt{a_{2-}a_{1-}}}\left(\int_{-e^{2}}^{e}\frac{dw}{(1-w^{2})\sqrt{\left[d^{2}-w^{2}\right]\left[e^{2}-w^{2}\right]}}+\int_{e^{2}}^{e}\frac{dw}{(1-w^{2})\sqrt{\left[d^{2}-w^{2}\right]\left[e^{2}-w^{2}\right]}}\right)
\]
Simplifying and regrouping, we obtain the expression for H, see Eq.
(\ref{eq:H}) as,

\begin{equation}
H=(a^{2}-b^{2}t_{+})I+\frac{{4b^{2}}}{\sqrt{{a_{2+}a_{1-}}}}\Pi(e^{2},m)\label{eq:finalH}
\end{equation}
where $\Pi$ is the complete elliptic integral of the third kind,
with $e^{2}<1$. This formula has been verified numerically. Also
in the above, we made use of the following step which was described
earlier in the previous section. It is as follows,

\[
\int_{-e^{2}}^{e}+\int_{e^{2}}^{e}=\int_{-e^{2}}^{0}+2\int_{0}^{e}-\int_{0}^{e^{2}}
\]
Then substituting $t=w/e$ to get the final form (\ref{eq:finalH}).
The complete elliptic integral of the third kind is given by,

\[
\Pi(e^{2},m)=\int_{0}^{1}\frac{{dt}}{(1-e^{2}t)\sqrt{(1-t^{2})(1-mt^{2})}}.
\]

\bibliographystyle{unsrt}
\bibliography{iRFP}

\end{document}